\documentclass[11pt]{article}

\usepackage{xurl}
\usepackage[hidelinks,breaklinks=true]{hyperref}
\usepackage{placeins}
\usepackage{amsmath, amsthm, amssymb}
\usepackage{graphicx}
\usepackage{geometry}
\usepackage{algorithm}
\usepackage{algpseudocode}
\usepackage[numbers]{natbib}
\newtheorem{definition}{Definition}

\newtheorem{theorem}{Theorem}

\usepackage{tikz}
\usepackage{float}
\usetikzlibrary{arrows.meta,positioning,calc}

\newcommand{\revcircleradius}{0.25}

\newcommand{\drawrevcircle}[1]{%
    \draw[draw=blue, line width=0.8pt] (#1) circle [radius=\revcircleradius];
}

\newcommand{\drawrevright}[1]{%
    \draw[->, line width=0.8pt, blue]
        ($(#1)+(0.22,-0.10)$) -- ($(#1)+(0.52,-0.40)$);
}

\newcommand{\drawrevleft}[1]{%
    \draw[->, line width=0.8pt, blue]
        ($(#1)+(-0.22,-0.10)$) -- ($(#1)+(-0.52,-0.40)$);
}

\title{
Strategic disclosure and epistemic manipulation in extensive games
}

\author{
\begin{tabular}{p{8cm}p{8cm}}
\centering Sergei Artemov & \centering Benjamin Prud'homme\tabularnewline
\centering \texttt{sartemov@gmail.com} &
\centering \texttt{benjamin.prudhomme46@login.cuny.edu}  \tabularnewline
\\
\multicolumn{2}{c}{CUNY Graduate Center} \\
\multicolumn{2}{c}{365 Fifth Avenue, New York City, NY 10016}
\end{tabular}
}

\date{\today}

\begin{document}

\maketitle

\begin{abstract}
In many strategic settings, agents can influence opponents by selectively disclosing truthful information about their own future actions. We introduce a game-theoretic framework to model this phenomenon. As a case study, we consider a two-player, alternating-move epistemic game in which Player $A$ evaluates actions under a conservative maximin rule, while Player $B$---the manipulator---leverages this behavior by strategically revealing commitments. We prove that the manipulator can implement every inducible outcome using at most $n$ disclosures in a binary game of depth $2n$ with $n$ decisions per player. This establishes a sharp contrast between the game tree's exponential size and the linear amount of information required to control the outcome, highlighting a general mechanism by which limited information can exert disproportionate strategic influence.
\end{abstract}

\section{Introduction}

In many strategic environments, agents must communicate truthfully about future behavior or relevant contingencies while retaining discretion over which commitments to disclose and when. Such selective disclosure can fundamentally alter an opponent's reasoning: by ruling out contingencies in advance, truthful commitments eliminate potential subgames, thereby reshaping the set of actions that appear rational to an observer.

For instance, legal and financial regulations often require agents to disclose accurate information about future actions or risks while allowing strategic choice over the timing and emphasis of such disclosures. Similarly, in bargaining or market interactions, a party may truthfully commit to a future pricing rule or reservation threshold, thereby influencing the opponent's planning. Other settings, such as public policy or industrial competition, exhibit the same underlying mechanism.

Standard solution concepts often presuppose common knowledge of rationality and rely on extensive updates that may not reflect real-world caution. To offer an alternative to this conventional doctrine, we introduce a general game-theoretic framework for studying epistemic manipulation under weaker epistemic assumptions. The core architecture of this framework rests on two main principles:

\begin{enumerate}
\item The manipulator exploits knowledge of opponents' preferences and specific epistemic conditions at relevant nodes to systematically alter their optimal responses.
\item The manipulation is executed via the truthful disclosure of committed moves at strategic nodes, pruning the game tree to alter the opponents' epistemic situation and subsequent behavior.
\end{enumerate}

As a case study, we apply this framework to two-player, alternating-move extensive-form games featuring asymmetric knowledge of epistemic states. Player $A$, lacking reliable a priori knowledge about $B$'s rationality, evaluates decisions conservatively using guaranteed maximin payoffs. Conversely, Player $B$---the manipulator---fully anticipates $A$'s cautious decision rule and leverages it by strategically disclosing truthful commitments about future actions. By isolating the recursive structure of conservative reasoning and the pruning effect of such commitments, we obtain precise bounds on the informational effort required to induce desired outcomes.

We prove that the manipulator can implement every inducible outcome using at most $n$ disclosures in a binary game of depth $2n$. This highlights a general mechanism by which limited information can exert disproportionate strategic influence: even in environments with exponentially many contingencies, a player can steer outcomes using only a linear number of commitments. This separation between combinatorial complexity and informational effort highlights a distinct mechanism of influence. 

Unlike models based on belief updating, where influence depends on shaping posterior beliefs, here influence operates by eliminating critical contingencies from consideration. More broadly, this reflects a general feature of decision-making under uncertainty: whenever actions are evaluated relative to a set of possible outcomes, selectively ruling out key contingencies can exert a disproportionate influence on behavior. While our analysis focuses on conservative (maximin) reasoning for clarity, the underlying mechanism is not tied to a specific decision rule; it points to a broader phenomenon that may persist in settings where choices depend on such sets more generally.

The remainder of the paper is organized as follows. Section 2 situates our approach within epistemic models of games, strategic information revelation, and information transmission. Section 3 develops intuition through simple examples illustrating recursive reasoning and the pruning effect of truthful commitments. Section 4 presents the main result, showing that inducible outcomes correspond to feasible paths from $A$'s perspective and can be implemented using at most $n$ revelations. Section 5 discusses computational and modeling extensions, including the search for optimal paths, dynamic revelation, and generalizations to richer game settings.

\section{Related Work}

\subsection{Epistemic Foundations and Knowledge-Based Rationality}

Classical decision-theoretic foundations model rational behavior via principles such as maximin reasoning, formalized by Harsanyi \cite{harsanyi_rational_1977} and rooted in earlier work by von Neumann and Morgenstern \cite{von_neumann_theory_2007}.

Our model builds on epistemic approaches to rational decision making in games. In the framework of knowledge-based rationality (KBR) introduced by Artemov~\cite{artemov_tr-2009015_2009}, players evaluate actions based on outcomes that are epistemically possible given their knowledge, selecting actions according to the highest known payoff. This framework departs from classical backward induction by allowing reasoning to depend explicitly on epistemic conditions. Artemov's epistemic game constructions, including examples in \cite{artemov_rational_2009}, illustrate how knowledge constraints affect strategic outcomes. In particular, they show that players' decisions may depend on what is known about future actions and about other players' rationality, highlighting the importance of epistemic structure in determining behavior. We return to this idea in Section 3 when discussing strategic information revelation.

These ideas are closely related to classical results on knowledge and rationality
in games. Under strong assumptions such as common knowledge of rationality,
backward induction emerges as the unique solution concept~\cite{aumann_backward_1995},
while earlier work highlights the role of shared knowledge in shaping agreement
and reasoning~\cite{aumann_agreeing_1976}. Our framework departs from these approaches by relaxing the assumption of common knowledge of rationality and allowing epistemic asymmetries to arise endogenously through strategic disclosure, rather than assuming a fixed epistemic structure.

\subsection{Epistemic Manipulation in Games}

The strategic role of information revelation has been studied within epistemic models of games. In Artemov \cite{artemov_tr-2009015_2009}, one player influences another's behavior by revealing truthful information about future actions or about other players' rationality, thereby altering the set of outcomes considered possible under conservative reasoning. A key insight from this framework is that increased knowledge can improve a player's guaranteed (known) payoff without necessarily improving their realized payoff.

Related observations appear in earlier work. Kamien, Tauman, and Zamir~\cite{kamien_value_1990} study the strategic value of information in games through an informed agent (a ``maven") who influences outcomes by selectively sharing information. Gerbrandy~\cite{gerbrandy_communication_2007} shows that partial disclosure can strictly outperform full disclosure, highlighting that the strategic value of information lies not only in its content, but in how it is revealed.

Parikh, Ta\c{s}demir, and Witzel~\cite{parikh_power_2013} build on this perspective by introducing a framework in which a knowledge manipulator (KM) influences outcomes by shaping the knowledge states of the active players. The KM does not act as a strategic player within the game, but instead operates externally, determining what players know or consider possible. While their framework can be reduced to a standard game in which the knowledge manipulator is treated as an additional player, they explicitly avoid this formulation, emphasizing instead the external role of information control.

Building on Artemov's framework and related perspectives on epistemic manipulation, we specialize to a two-player, alternating-move setting while allowing for multiple levels per player, yielding a tractable model that captures the recursive structure of strategic reasoning. Within this setting, our model develops a more structured, path-based analysis of strategic information revelation. In particular, we make explicit how truthful commitments prune feasible continuations from the game tree and thereby reshape the opponent's recursive reasoning process. This perspective allows us to analyze how revelation interacts with incentives and to derive structural and quantitative bounds on the amount of information required to induce specific outcomes. Unlike prior approaches that treat information control as external to the game or emphasize belief updating as the primary mechanism, our framework embeds truthful revelation directly as a strategic action within the game.
\subsection{Information Transmission and Communication}

A related literature studies how information transmission affects strategic behavior. In Bayesian persuasion~\cite{kamenica_bayesian_2011} and, more broadly, information design~\cite{bergemann_information_2019}, a sender influences a receiver's action by shaping posterior beliefs through signals. Similarly, in cheap talk models~\cite{crawford_strategic_1982}, communication occurs through costless, non-binding messages that affect beliefs in equilibrium.

These approaches emphasize belief updating as the primary mechanism through which information affects decisions. Formal semantic frameworks, such as that of Parikh and Ramanujam\cite{parikh_knowledge_2003}, provide tools for reasoning about the informational content of messages and their effects on knowledge. In contrast to these approaches, our framework does not model information transmission through belief updating. Instead, truthful revelation prunes feasible continuations from the game tree, directly shaping the opponent's recursive reasoning.

\subsection{Discussion and Positioning}

Taken together, these perspectives motivate an approach in which epistemic manipulation is embedded directly within the strategic interaction. Rather than introducing an external manipulator or relying on belief updates, a player strategically reveals truthful information about their own future actions, thereby restricting the set of feasible continuations considered by the opponent.

This endogenous formulation has three key consequences. First, manipulation is itself a strategic action subject to incentives and constraints within the game. Second, information affects behavior by pruning feasible outcomes rather than by updating beliefs. Third, the framework enables a quantitative analysis of informational efficiency, allowing us to characterize how much information is required to induce a desired outcome.

Our contribution is to integrate epistemic reasoning and strategic information revelation into a unified framework and to establish structural bounds on the amount of information required to induce a target outcome.

Conceptually, this framework enables quantitative analysis of epistemic manipulation. In particular, our basic model describes a meaningful scenario in which a non-expensive (linear in the depth of the game) manipulation is sufficient to alter the exponential-size game to take any theoretically possible path.

\section{Intuition: A simple motivating example}

In this section we build intuition through small examples showing how player $A$'s conservative reasoning
interacts with information revealed by player $B$, and how such
revelations can influence the set of outcomes that $A$ considers
possible.

\subsection{The one-level-per-player case}

First let us consider the simplest setting, in which each player moves
exactly once.

\begin{figure}[H]
\centering
\begin{tikzpicture}[
    x=1cm, y=1cm,
    >={Stealth[length=3mm,width=2mm]},
    every node/.style={inner sep=0pt},
    player/.style={font=\Large},
    payoff/.style={font=\Large},
    minlabel/.style={font=\large\color{green!50!black}},
    node dot/.style={circle, fill=black, inner sep=1.6pt},
    black edge/.style={->, line width=0.4pt, black},
    red edge/.style={->, line width=0.8pt, red}
]

\coordinate (A)  at (0,4.2);

\coordinate (BL) at (-4.6,2.2);
\coordinate (BR) at ( 4.6,2.2);

\coordinate (L1) at (-6.0,0.0);
\coordinate (L2) at (-3.2,0.0);
\coordinate (R1) at ( 3.2,0.0);
\coordinate (R2) at ( 6.0,0.0);

\node[node dot] (Adot)  at (A)  {};
\node[node dot] (BLdot) at (BL) {};
\node[node dot] (BRdot) at (BR) {};

\node[player] at ($(A)+(0.35,0.22)$) {$A$};
\node[player] at ($(BL)+(-0.55,0.10)$) {$B$};
\node[player] at ($(BR)+(0.55,0.10)$) {$B$};

\node[payoff] (L1lab) at (L1) {$\binom{0}{1}$};
\node[payoff] (L2lab) at (L2) {$\binom{2}{3}$};
\node[payoff] (R1lab) at (R1) {$\binom{1}{0}$};
\node[payoff] (R2lab) at (R2) {$\binom{3}{2}$};

\draw[black edge] (Adot)  -- (BLdot);
\draw[black edge] (Adot)  -- (BRdot);

\draw[black edge] (BLdot) -- ($(L1lab)+(0,0.45)$);
\draw[black edge] (BLdot) -- ($(L2lab)+(0,0.45)$);
\draw[black edge] (BRdot) -- ($(R1lab)+(0,0.45)$);
\draw[black edge] (BRdot) -- ($(R2lab)+(0,0.45)$);

\node[minlabel] at ($(A)!0.53!(BL)+(-0.25,0.45)$) {$\min\ 0$};
\node[minlabel] at ($(A)!0.53!(BR)+(0.25,0.45)$) {$\min\ 1$};

\draw[red edge] (Adot) -- (BRdot);
\draw[red edge] (BRdot) -- ($(R2lab)+(0,0.45)$);

\end{tikzpicture}
\caption{A one-level-per-player game. The green $\min$ labels indicate the guaranteed payoff of each action for player $A$, computed as the minimum payoff among outcomes reachable along that branch. The red arrows indicate the path of play under the players' decision rules.}
\end{figure}
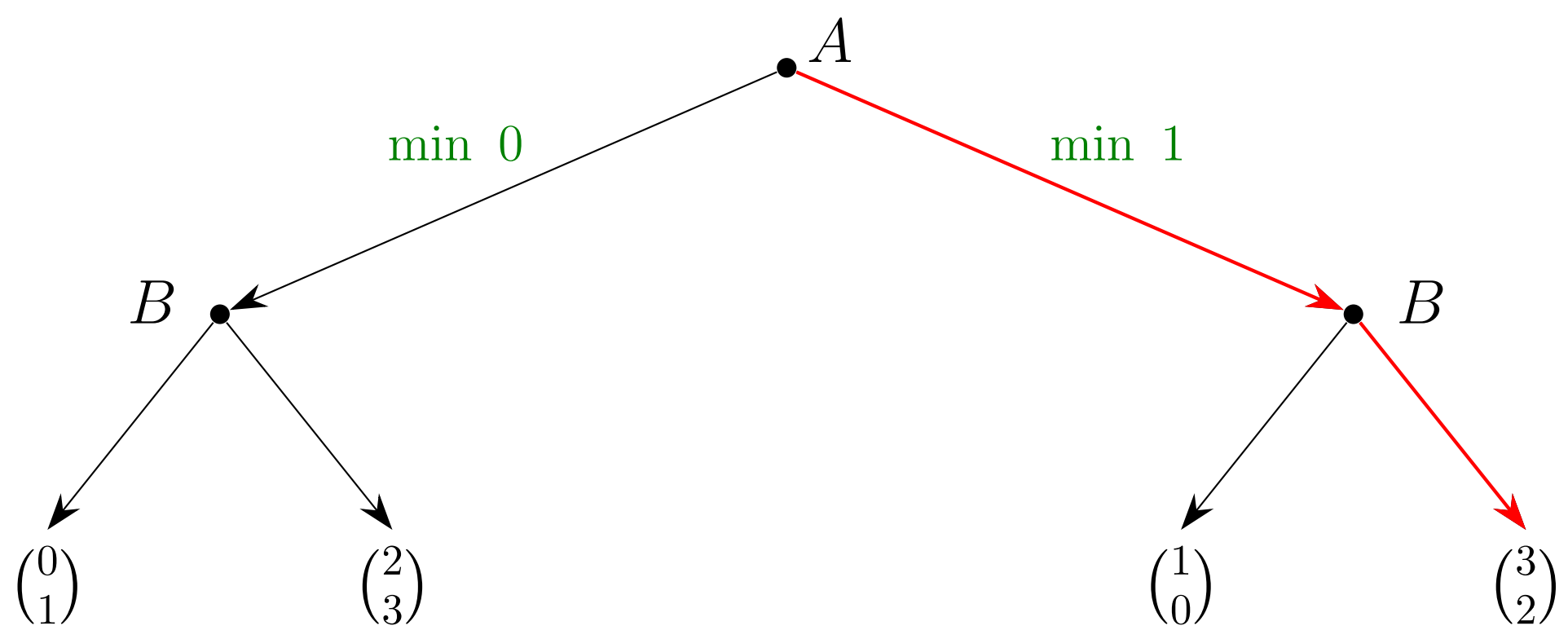

From the root (Figure 1), both available directions lead to two possible outcomes. Because player $A$ cannot predict $B$'s behavior beyond the structure of the game tree, they evaluate each direction conservatively by its \emph{guaranteed payoff}, defined as the minimum payoff to $A$ among the outcomes that remain possible if that direction is chosen. In contrast, the \emph{real payoff} is the payoff actually obtained once player $B$ selects an action. This distinction is what makes $A$'s decisions sensitive to which outcomes are considered possible, and hence to information revealed by $B$.

Here, if $A$ moves left, the possible outcomes yield payoffs $0$ and $2$, so the guaranteed payoff is $0$. If $A$ moves right, the outcomes yield payoffs $1$ and $3$, giving a guaranteed payoff of $1$. Thus $A$ prefers to move right, and player $B$ then chooses the action maximizing their own payoff, leading to the outcome $(3,2)$. If $A$ were instead to move left, player $B$ would choose the action leading to payoff $2$, so the real payoff would be $2$.

Note that the comparison of $A$'s guaranteed payoffs relies on the fact that the guaranteed payoffs are distinct. 
Throughout, we restrict attention to generic games, in which no two terminal outcomes 
give the same payoff to a given player. This ensures that all such comparisons are 
strict, so player $A$'s choices are uniquely determined.

\begin{figure}[H]
\centering
\begin{tikzpicture}[
    x=1cm, y=1cm,
    >={Stealth[length=3mm,width=2mm]},
    every node/.style={inner sep=0pt},
    player/.style={font=\Large},
    payoff/.style={font=\Large},
    minlabel/.style={font=\large\color{green!50!black}},
    node dot/.style={circle, fill=black, inner sep=1.6pt},
    black edge/.style={->, line width=0.4pt, black},
    red edge/.style={->, line width=0.8pt, red},
    blue ann/.style={->, line width=0.8pt, blue!70!black},
    blue circ/.style={draw=blue!70!black, line width=0.8pt}
]

\coordinate (A)  at (0,4.2);

\coordinate (BL) at (-4.6,2.2);
\coordinate (BR) at ( 4.6,2.2);

\coordinate (L1) at (-6.0,0.0);
\coordinate (L2) at (-3.2,0.0);
\coordinate (R1) at ( 3.2,0.0);
\coordinate (R2) at ( 6.0,0.0);

\node[node dot] (Adot)  at (A)  {};
\node[node dot] (BLdot) at (BL) {};
\node[node dot] (BRdot) at (BR) {};

\node[player] at ($(A)+(0.35,0.22)$) {$A$};
\node[player] at ($(BL)+(-0.55,0.10)$) {$B$};
\node[player] at ($(BR)+(0.55,0.10)$) {$B$};

\node[payoff] (L1lab) at (L1) {$\binom{0}{1}$};
\node[payoff] (L2lab) at (L2) {$\binom{2}{3}$};
\node[payoff] (R1lab) at (R1) {$\binom{1}{0}$};
\node[payoff] (R2lab) at (R2) {$\binom{3}{2}$};

\draw[black edge] (Adot)  -- (BLdot);
\draw[black edge] (Adot)  -- (BRdot);

\draw[black edge] (BLdot) -- ($(L1lab)+(0,0.45)$);
\draw[black edge] (BLdot) -- ($(L2lab)+(0,0.45)$);
\draw[black edge] (BRdot) -- ($(R1lab)+(0,0.45)$);
\draw[black edge] (BRdot) -- ($(R2lab)+(0,0.45)$);

\node[minlabel] at ($(A)!0.53!(BL)+(-0.25,0.45)$) {$\min\ 2$};
\node[minlabel] at ($(A)!0.53!(BR)+(0.25,0.45)$) {$\min\ 1$};

\draw[red edge] (Adot) -- (BLdot);
\draw[red edge] (BLdot) -- ($(L2lab)+(0,0.45)$);

\drawrevcircle{BLdot}
\drawrevright{BLdot}

\draw[red, line width=0.9pt]
    ($(L1lab)+(-0.28,0.28)$) -- ($(L1lab)+(0.28,-0.28)$);
\draw[red, line width=0.9pt]
    ($(L1lab)+(-0.28,-0.28)$) -- ($(L1lab)+(0.28,0.28)$);

\end{tikzpicture}
\caption{A revelation that sways player $A$'s decision. The blue circle and arrow indicate player $B$'s truthful commitment to choose the right-hand outcome if the left branch is reached. The red X marks the outcome thereby ruled out from player $A$'s consideration. The green $\min$ labels show the resulting guaranteed payoffs for player $A$, and the red arrows indicate the induced path of play.}
\end{figure}
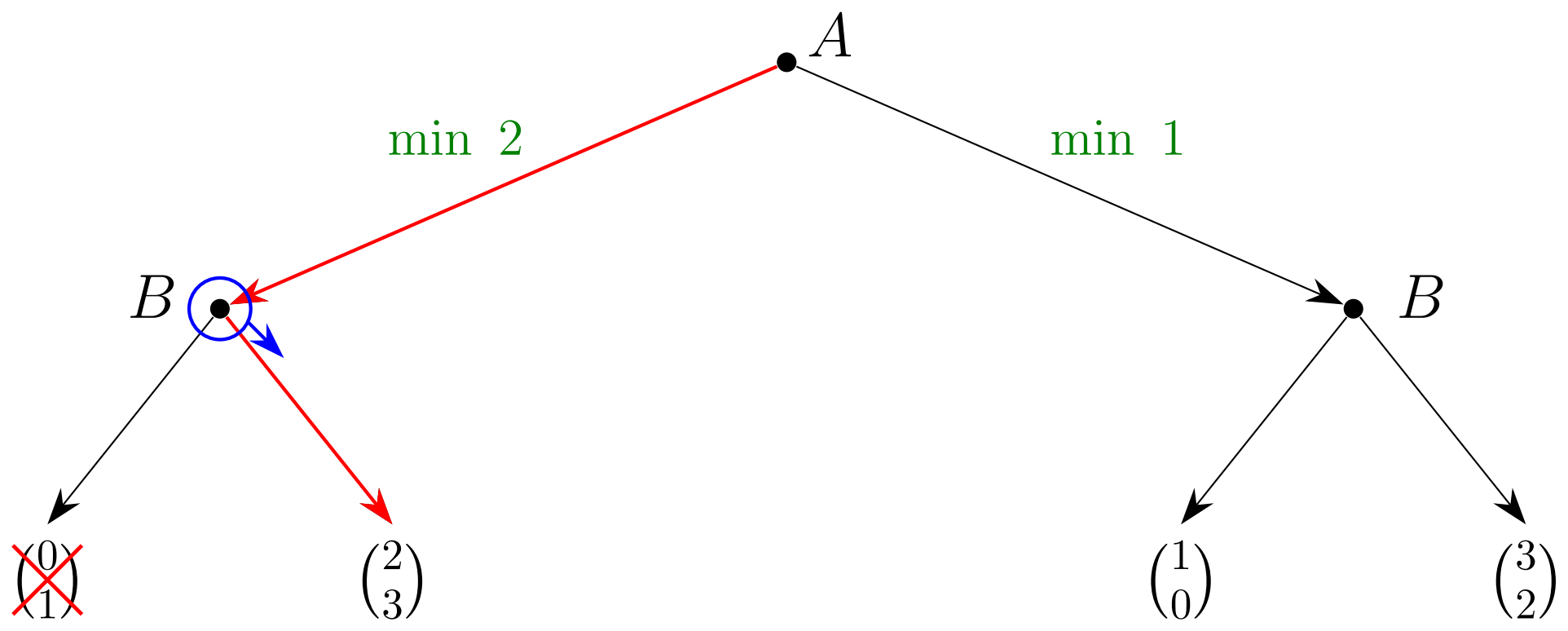

Suppose now (Figure 2) that player $B$ reveals that, if the left branch is reached, they will choose the action leading to the right-hand outcome. From $A$'s perspective, the leftmost outcome is therefore no longer feasible and can be pruned from consideration. The remaining outcome has payoff 2, so the guaranteed payoff of moving left increases from 0 to 2, exceeding that of moving right. Consequently, $A$ is swayed to move left. This change benefits player $B$: without revealing information, $B$ would receive payoff 2 (since $A$ moves right), whereas after the revelation, $A$ moves left and $B$ obtains payoff 3.

More generally, since $A$ compares guaranteed payoffs when choosing actions, only revelations that raise the smaller guaranteed payoff of one direction above that of the other can influence $A$'s decision. Player $B$ is therefore motivated to reveal information that steers $A$ toward paths yielding higher payoffs for $B$.

\begin{figure}[H]
\centering
\begin{tikzpicture}[
    x=1cm, y=1cm,
    >={Stealth[length=3mm,width=2mm]},
    every node/.style={inner sep=0pt},
    player/.style={font=\Large},
    payoff/.style={font=\Large},
    minlabel/.style={font=\large\color{green!50!black}},
    node dot/.style={circle, fill=black, inner sep=1.6pt},
    black edge/.style={->, line width=0.4pt, black},
    red edge/.style={->, line width=0.8pt, red},
    blue ann/.style={->, line width=0.8pt, blue!70!black},
    blue circ/.style={draw=blue!70!black, line width=0.8pt}
]

\coordinate (A)  at (0,4.2);

\coordinate (BL) at (-4.6,2.2);
\coordinate (BR) at ( 4.6,2.2);

\coordinate (L1) at (-6.0,0.0);
\coordinate (L2) at (-3.2,0.0);
\coordinate (R1) at ( 3.2,0.0);
\coordinate (R2) at ( 6.0,0.0);

\node[node dot] (Adot)  at (A)  {};
\node[node dot] (BLdot) at (BL) {};
\node[node dot] (BRdot) at (BR) {};

\node[player] at ($(A)+(0.35,0.22)$) {$A$};
\node[player] at ($(BL)+(-0.55,0.10)$) {$B$};
\node[player] at ($(BR)+(0.55,0.10)$) {$B$};

\node[payoff] (L1lab) at (L1) {$\binom{\textcolor{red}{2}}{1}$};
\node[payoff] (L2lab) at (L2) {$\binom{\textcolor{red}{0}}{3}$};
\node[payoff] (R1lab) at (R1) {$\binom{1}{0}$};
\node[payoff] (R2lab) at (R2) {$\binom{3}{2}$};

\draw[black edge] (Adot)  -- (BLdot);
\draw[black edge] (Adot)  -- (BRdot);

\draw[black edge] (BLdot) -- ($(L1lab)+(0,0.45)$);
\draw[black edge] (BLdot) -- ($(L2lab)+(0,0.45)$);
\draw[black edge] (BRdot) -- ($(R1lab)+(0,0.45)$);
\draw[black edge] (BRdot) -- ($(R2lab)+(0,0.45)$);

\node[minlabel] at ($(A)!0.53!(BL)+(-0.25,0.45)$) {$\min\ 0$};
\node[minlabel] at ($(A)!0.53!(BR)+(0.25,0.45)$) {$\min\ 1$};

\draw[red edge] (Adot) -- (BRdot);
\draw[red edge] (BRdot) -- ($(R2lab)+(0,0.45)$);

\drawrevcircle{BLdot}
\drawrevright{BLdot}

\draw[red, line width=0.9pt]
    ($(L1lab)+(-0.28,0.28)$) -- ($(L1lab)+(0.28,-0.28)$);
\draw[red, line width=0.9pt]
    ($(L1lab)+(-0.28,-0.28)$) -- ($(L1lab)+(0.28,0.28)$);

\end{tikzpicture}
\caption{A revelation that fails to sway player $A$. Relative to Figure~2, the red-colored payoffs on the two leftmost outcomes are interchanged. With this change, the same truthful revelation by player $B$ no longer raises the guaranteed payoff of moving left above that of moving right, so player $A$ continues to choose the right branch.}
\end{figure}
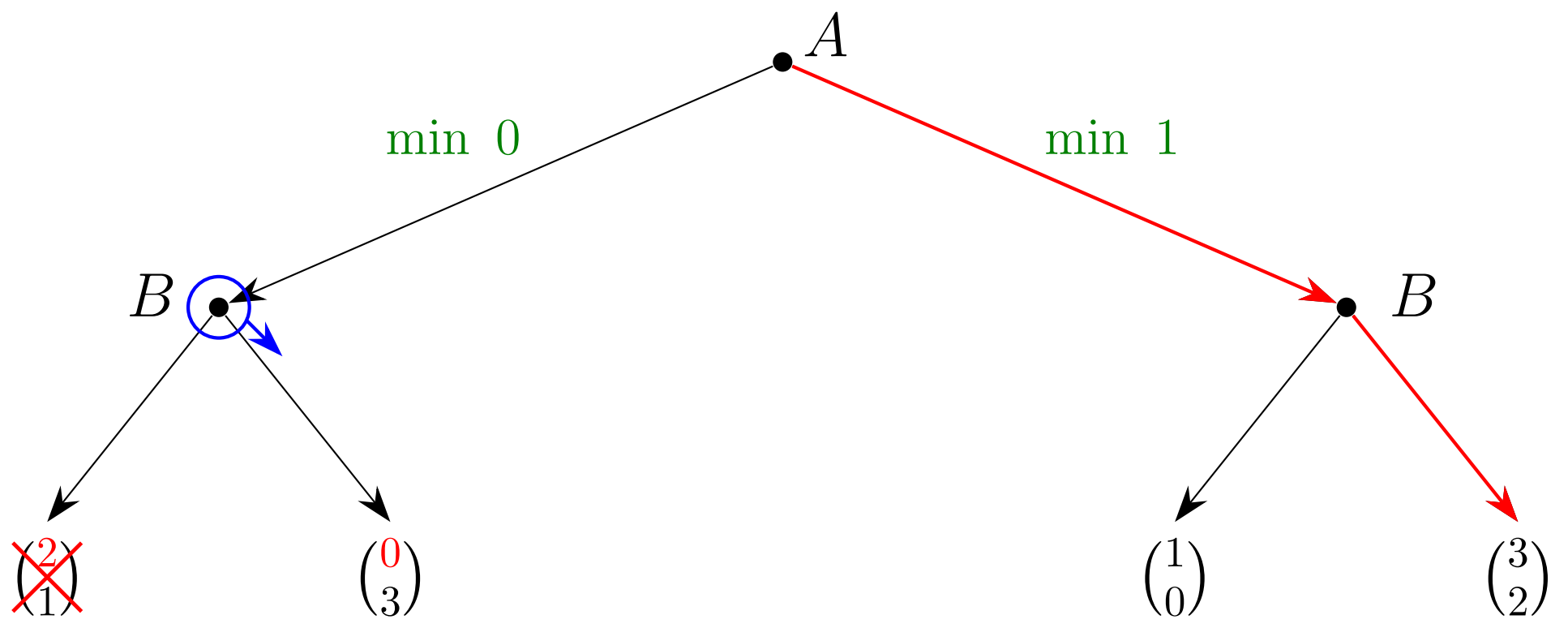

However, such manipulation is not always possible. Consider the same tree with the payoffs of the two left outcomes interchanged (Figure 3). In this case, even after $B$ reveals their intended move at the left node, the outcome that is pruned is not the one determining the minimum payoff for $A$. The guaranteed payoff of moving left therefore remains lower than that of moving right, and $A$ continues to choose the right branch. We say that, in this configuration, neither of the outcomes reachable from the left branch is \emph{inducible}.

The simplicity of the one-level-per-player case stems from the fact that $A$ makes no further decisions after the root. In deeper games, however, $A$ must reason about future decision nodes, and this recursive structure fundamentally changes how guaranteed payoffs are computed.

\subsection{The two-level-per-player case}

We now consider a game in which each player moves twice. This can be viewed as four one-level-per-player subgames beneath an additional decision level for each player (Figure 4).

\begin{figure}[H]
\centering
\resizebox{\textwidth}{!}{%
\begin{tikzpicture}[
    x=1cm, y=1cm,
    >={Stealth[length=3mm,width=2mm]},
    every node/.style={inner sep=0pt},
    player/.style={font=\Large},
    payoff/.style={font=\Large},
    node dot/.style={circle, fill=black, inner sep=1.6pt},
    black edge/.style={->, line width=0.4pt, black}
]

\coordinate (A0) at (0,6.0);

\coordinate (B1L) at (-4.8,4.5);
\coordinate (B1R) at ( 4.8,4.5);

\coordinate (A2LL) at (-7.2,3.0);
\coordinate (A2LR) at (-2.4,3.0);
\coordinate (A2RL) at ( 2.4,3.0);
\coordinate (A2RR) at ( 7.2,3.0);

\coordinate (B3LLL) at (-8.4,1.5);
\coordinate (B3LLR) at (-6.0,1.5);

\coordinate (B3LRL) at (-3.6,1.5);
\coordinate (B3LRR) at (-1.2,1.5);

\coordinate (B3RLL) at ( 1.2,1.5);
\coordinate (B3RLR) at ( 3.6,1.5);

\coordinate (B3RRL) at ( 6.0,1.5);
\coordinate (B3RRR) at ( 8.4,1.5);

\coordinate (L1)  at (-9.0,0.0);
\coordinate (L2)  at (-7.8,0.0);
\coordinate (L3)  at (-6.6,0.0);
\coordinate (L4)  at (-5.4,0.0);

\coordinate (L5)  at (-4.2,0.0);
\coordinate (L6)  at (-3.0,0.0);
\coordinate (L7)  at (-1.8,0.0);
\coordinate (L8)  at (-0.6,0.0);

\coordinate (L9)  at ( 0.6,0.0);
\coordinate (L10) at ( 1.8,0.0);
\coordinate (L11) at ( 3.0,0.0);
\coordinate (L12) at ( 4.2,0.0);

\coordinate (L13) at ( 5.4,0.0);
\coordinate (L14) at ( 6.6,0.0);
\coordinate (L15) at ( 7.8,0.0);
\coordinate (L16) at ( 9.0,0.0);

\node[node dot] (A0dot)    at (A0)    {};
\node[node dot] (B1Ldot)   at (B1L)   {};
\node[node dot] (B1Rdot)   at (B1R)   {};

\node[node dot] (A2LLdot)  at (A2LL)  {};
\node[node dot] (A2LRdot)  at (A2LR)  {};
\node[node dot] (A2RLdot)  at (A2RL)  {};
\node[node dot] (A2RRdot)  at (A2RR)  {};

\node[node dot] (B3LLLdot) at (B3LLL) {};
\node[node dot] (B3LLRdot) at (B3LLR) {};
\node[node dot] (B3LRLdot) at (B3LRL) {};
\node[node dot] (B3LRRdot) at (B3LRR) {};
\node[node dot] (B3RLLdot) at (B3RLL) {};
\node[node dot] (B3RLRdot) at (B3RLR) {};
\node[node dot] (B3RRLdot) at (B3RRL) {};
\node[node dot] (B3RRRdot) at (B3RRR) {};

\node[player] at ($(A0)+(0.35,0.22)$) {$A$};

\node[player] at ($(B1L)+(-0.55,0.10)$) {$B$};
\node[player] at ($(B1R)+(0.55,0.10)$)  {$B$};

\node[player] at ($(A2LL)+(-0.45,0.10)$) {$A$};
\node[player] at ($(A2LR)+(0.45,0.10)$) {$A$};\node[player] at ($(A2RL)+(-0.45,0.10)$) {$A$};
\node[player] at ($(A2RR)+(0.45,0.10)$) {$A$};

\node[player] at ($(B3LLL)+(-0.35,0.10)$) {$B$};
\node[player] at ($(B3LLR)+(0.35,0.10)$)  {$B$};
\node[player] at ($(B3LRL)+(-0.35,0.10)$) {$B$};
\node[player] at ($(B3LRR)+(0.35,0.10)$)  {$B$};
\node[player] at ($(B3RLL)+(-0.35,0.10)$) {$B$};
\node[player] at ($(B3RLR)+(0.35,0.10)$)  {$B$};
\node[player] at ($(B3RRL)+(-0.35,0.10)$) {$B$};
\node[player] at ($(B3RRR)+(0.35,0.10)$)  {$B$};

\node[payoff] (L1lab)  at (L1)  {$\binom{9}{15}$};
\node[payoff] (L2lab)  at (L2)  {$\binom{2}{4}$};
\node[payoff] (L3lab)  at (L3)  {$\binom{10}{0}$};
\node[payoff] (L4lab)  at (L4)  {$\binom{7}{10}$};

\node[payoff] (L5lab)  at (L5)  {$\binom{1}{7}$};
\node[payoff] (L6lab)  at (L6)  {$\binom{11}{9}$};
\node[payoff] (L7lab)  at (L7)  {$\binom{4}{2}$};
\node[payoff] (L8lab)  at (L8)  {$\binom{5}{3}$};

\node[payoff] (L9lab)  at (L9)  {$\binom{15}{1}$};
\node[payoff] (L10lab) at (L10) {$\binom{0}{12}$};
\node[payoff] (L11lab) at (L11) {$\binom{8}{14}$};
\node[payoff] (L12lab) at (L12) {$\binom{13}{8}$};

\node[payoff] (L13lab) at (L13) {$\binom{3}{13}$};
\node[payoff] (L14lab) at (L14) {$\binom{14}{5}$};
\node[payoff] (L15lab) at (L15) {$\binom{6}{11}$};
\node[payoff] (L16lab) at (L16) {$\binom{12}{6}$};

\draw[black edge] (A0dot) -- (B1Ldot);
\draw[black edge] (A0dot) -- (B1Rdot);

\draw[black edge] (B1Ldot) -- (A2LLdot);
\draw[black edge] (B1Ldot) -- (A2LRdot);
\draw[black edge] (B1Rdot) -- (A2RLdot);
\draw[black edge] (B1Rdot) -- (A2RRdot);

\draw[black edge] (A2LLdot) -- (B3LLLdot);
\draw[black edge] (A2LLdot) -- (B3LLRdot);
\draw[black edge] (A2LRdot) -- (B3LRLdot);
\draw[black edge] (A2LRdot) -- (B3LRRdot);
\draw[black edge] (A2RLdot) -- (B3RLLdot);
\draw[black edge] (A2RLdot) -- (B3RLRdot);
\draw[black edge] (A2RRdot) -- (B3RRLdot);
\draw[black edge] (A2RRdot) -- (B3RRRdot);

\draw[black edge] (B3LLLdot) -- ($(L1lab)+(0,0.45)$);
\draw[black edge] (B3LLLdot) -- ($(L2lab)+(0,0.45)$);
\draw[black edge] (B3LLRdot) -- ($(L3lab)+(0,0.45)$);
\draw[black edge] (B3LLRdot) -- ($(L4lab)+(0,0.45)$);

\draw[black edge] (B3LRLdot) -- ($(L5lab)+(0,0.45)$);
\draw[black edge] (B3LRLdot) -- ($(L6lab)+(0,0.45)$);
\draw[black edge] (B3LRRdot) -- ($(L7lab)+(0,0.45)$);
\draw[black edge] (B3LRRdot) -- ($(L8lab)+(0,0.45)$);

\draw[black edge] (B3RLLdot) -- ($(L9lab)+(0,0.45)$);
\draw[black edge] (B3RLLdot) -- ($(L10lab)+(0,0.45)$);
\draw[black edge] (B3RLRdot) -- ($(L11lab)+(0,0.45)$);
\draw[black edge] (B3RLRdot) -- ($(L12lab)+(0,0.45)$);

\draw[black edge] (B3RRLdot) -- ($(L13lab)+(0,0.45)$);
\draw[black edge] (B3RRLdot) -- ($(L14lab)+(0,0.45)$);
\draw[black edge] (B3RRRdot) -- ($(L15lab)+(0,0.45)$);
\draw[black edge] (B3RRRdot) -- ($(L16lab)+(0,0.45)$);

\end{tikzpicture}
}
\caption{A two-level-per-player game.}
\end{figure}
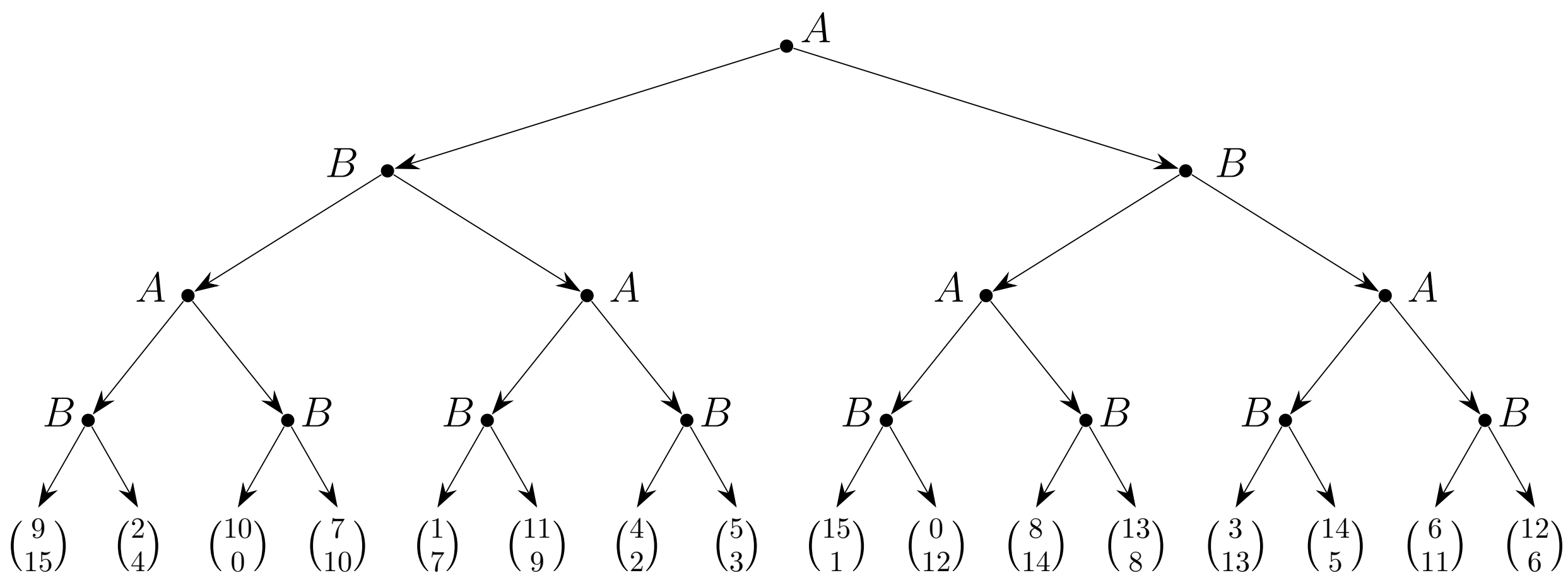

In this deeper tree, determining optimal behavior requires recursive reasoning. At higher-level nodes, $A$ computes guaranteed payoffs recursively by anticipating their own future choices, while $B$ anticipates the resulting outcomes under optimal play.

\medskip
\noindent
\textbf{Level-2 reasoning.}
At level-2 $B$-nodes, a move leads directly to a terminal outcome, so $B$ selects the direction yielding the higher payoff, determining a unique realized outcome at each such node. At each level-2 $A$-node, player $A$ computes the guaranteed payoff of each direction exactly as in the one-level-per-player case: for each direction, this is the minimum of $A$'s payoffs among the reachable terminal outcomes. Player $A$ then selects the direction with the higher guaranteed payoff. Thus, after level-2 reasoning, each such node is associated with both a realized outcome and a set of outcomes that remain possible for $A$ under conservative reasoning (Figure 5).

\begin{figure}[H]
\centering
\resizebox{\textwidth}{!}{%
\begin{tikzpicture}[
    x=1cm, y=1cm,
    >={Stealth[length=3mm,width=2mm]},
    every node/.style={inner sep=0pt},
    player/.style={font=\Large},
    payoff/.style={font=\Large},
    minlabel/.style={font=\large\color{green!50!black}},
    node dot/.style={circle, fill=black, inner sep=1.6pt},
    black edge/.style={->, line width=0.4pt, black},
    red edge/.style={->, line width=0.8pt, red},
    green ul/.style={line width=1.1pt, color=green!50!black},
    red ul/.style={line width=1.1pt, color=red!80!black}
]

\coordinate (A0) at (0,6.0);

\coordinate (B1L) at (-4.8,4.5);
\coordinate (B1R) at ( 4.8,4.5);

\coordinate (A2LL) at (-7.2,3.0);
\coordinate (A2LR) at (-2.4,3.0);
\coordinate (A2RL) at ( 2.4,3.0);
\coordinate (A2RR) at ( 7.2,3.0);

\coordinate (B3LLL) at (-8.4,1.5);
\coordinate (B3LLR) at (-6.0,1.5);

\coordinate (B3LRL) at (-3.6,1.5);
\coordinate (B3LRR) at (-1.2,1.5);

\coordinate (B3RLL) at ( 1.2,1.5);
\coordinate (B3RLR) at ( 3.6,1.5);

\coordinate (B3RRL) at ( 6.0,1.5);
\coordinate (B3RRR) at ( 8.4,1.5);

\coordinate (L1)  at (-9.0,0.0);
\coordinate (L2)  at (-7.8,0.0);
\coordinate (L3)  at (-6.6,0.0);
\coordinate (L4)  at (-5.4,0.0);

\coordinate (L5)  at (-4.2,0.0);
\coordinate (L6)  at (-3.0,0.0);
\coordinate (L7)  at (-1.8,0.0);
\coordinate (L8)  at (-0.6,0.0);

\coordinate (L9)  at ( 0.6,0.0);
\coordinate (L10) at ( 1.8,0.0);
\coordinate (L11) at ( 3.0,0.0);
\coordinate (L12) at ( 4.2,0.0);

\coordinate (L13) at ( 5.4,0.0);
\coordinate (L14) at ( 6.6,0.0);
\coordinate (L15) at ( 7.8,0.0);
\coordinate (L16) at ( 9.0,0.0);

\node[node dot] (A0dot)    at (A0)    {};
\node[node dot] (B1Ldot)   at (B1L)   {};
\node[node dot] (B1Rdot)   at (B1R)   {};

\node[node dot] (A2LLdot)  at (A2LL)  {};
\node[node dot] (A2LRdot)  at (A2LR)  {};
\node[node dot] (A2RLdot)  at (A2RL)  {};
\node[node dot] (A2RRdot)  at (A2RR)  {};

\node[node dot] (B3LLLdot) at (B3LLL) {};
\node[node dot] (B3LLRdot) at (B3LLR) {};
\node[node dot] (B3LRLdot) at (B3LRL) {};
\node[node dot] (B3LRRdot) at (B3LRR) {};
\node[node dot] (B3RLLdot) at (B3RLL) {};
\node[node dot] (B3RLRdot) at (B3RLR) {};
\node[node dot] (B3RRLdot) at (B3RRL) {};
\node[node dot] (B3RRRdot) at (B3RRR) {};

\node[player] at ($(A0)+(0.35,0.22)$) {$A$};

\node[player] at ($(B1L)+(-0.55,0.10)$) {$B$};
\node[player] at ($(B1R)+(0.55,0.10)$)  {$B$};

\node[player] at ($(A2LL)+(-0.45,0.10)$) {$A$};
\node[player] at ($(A2LR)+(0.45,0.10)$)  {$A$};
\node[player] at ($(A2RL)+(-0.45,0.10)$) {$A$};
\node[player] at ($(A2RR)+(0.45,0.10)$)  {$A$};

\node[player] at ($(B3LLL)+(-0.35,0.10)$) {$B$};
\node[player] at ($(B3LLR)+(0.35,0.10)$)  {$B$};
\node[player] at ($(B3LRL)+(-0.35,0.10)$) {$B$};
\node[player] at ($(B3LRR)+(0.35,0.10)$)  {$B$};
\node[player] at ($(B3RLL)+(-0.35,0.10)$) {$B$};
\node[player] at ($(B3RLR)+(0.35,0.10)$)  {$B$};
\node[player] at ($(B3RRL)+(-0.35,0.10)$) {$B$};
\node[player] at ($(B3RRR)+(0.35,0.10)$)  {$B$};

\node[payoff] (L1lab)  at (L1)  {$\binom{9}{15}$};
\node[payoff] (L2lab)  at (L2)  {$\binom{2}{4}$};
\node[payoff] (L3lab)  at (L3)  {$\binom{10}{0}$};
\node[payoff] (L4lab)  at (L4)  {$\binom{7}{10}$};

\node[payoff] (L5lab)  at (L5)  {$\binom{1}{7}$};
\node[payoff] (L6lab)  at (L6)  {$\binom{11}{9}$};
\node[payoff] (L7lab)  at (L7)  {$\binom{4}{2}$};
\node[payoff] (L8lab)  at (L8)  {$\binom{5}{3}$};

\node[payoff] (L9lab)  at (L9)  {$\binom{15}{1}$};
\node[payoff] (L10lab) at (L10) {$\binom{0}{12}$};
\node[payoff] (L11lab) at (L11) {$\binom{8}{14}$};
\node[payoff] (L12lab) at (L12) {$\binom{13}{8}$};

\node[payoff] (L13lab) at (L13) {$\binom{3}{13}$};
\node[payoff] (L14lab) at (L14) {$\binom{14}{5}$};
\node[payoff] (L15lab) at (L15) {$\binom{6}{11}$};
\node[payoff] (L16lab) at (L16) {$\binom{12}{6}$};

\draw[black edge] (A0dot) -- (B1Ldot);
\draw[black edge] (A0dot) -- (B1Rdot);

\draw[black edge] (B1Ldot) -- (A2LLdot);
\draw[black edge] (B1Ldot) -- (A2LRdot);
\draw[black edge] (B1Rdot) -- (A2RLdot);
\draw[black edge] (B1Rdot) -- (A2RRdot);

\draw[black edge] (A2LLdot) -- (B3LLLdot);
\draw[black edge] (A2LLdot) -- (B3LLRdot);
\draw[black edge] (A2LRdot) -- (B3LRLdot);
\draw[black edge] (A2LRdot) -- (B3LRRdot);
\draw[black edge] (A2RLdot) -- (B3RLLdot);
\draw[black edge] (A2RLdot) -- (B3RLRdot);
\draw[black edge] (A2RRdot) -- (B3RRLdot);
\draw[black edge] (A2RRdot) -- (B3RRRdot);

\draw[black edge] (B3LLLdot) -- ($(L1lab)+(0,0.45)$);
\draw[black edge] (B3LLLdot) -- ($(L2lab)+(0,0.45)$);
\draw[black edge] (B3LLRdot) -- ($(L3lab)+(0,0.45)$);
\draw[black edge] (B3LLRdot) -- ($(L4lab)+(0,0.45)$);

\draw[black edge] (B3LRLdot) -- ($(L5lab)+(0,0.45)$);
\draw[black edge] (B3LRLdot) -- ($(L6lab)+(0,0.45)$);
\draw[black edge] (B3LRRdot) -- ($(L7lab)+(0,0.45)$);
\draw[black edge] (B3LRRdot) -- ($(L8lab)+(0,0.45)$);

\draw[black edge] (B3RLLdot) -- ($(L9lab)+(0,0.45)$);
\draw[black edge] (B3RLLdot) -- ($(L10lab)+(0,0.45)$);
\draw[black edge] (B3RLRdot) -- ($(L11lab)+(0,0.45)$);
\draw[black edge] (B3RLRdot) -- ($(L12lab)+(0,0.45)$);

\draw[black edge] (B3RRLdot) -- ($(L13lab)+(0,0.45)$);
\draw[black edge] (B3RRLdot) -- ($(L14lab)+(0,0.45)$);
\draw[black edge] (B3RRRdot) -- ($(L15lab)+(0,0.45)$);
\draw[black edge] (B3RRRdot) -- ($(L16lab)+(0,0.45)$);

\draw[red edge] (A2LLdot) -- (B3LLRdot);
\draw[red edge] (A2LRdot) -- (B3LRRdot);
\draw[red edge] (A2RLdot) -- (B3RLRdot);
\draw[red edge] (A2RRdot) -- (B3RRRdot);

\draw[red edge] (B3LLLdot) -- ($(L1lab)+(0,0.45)$);
\draw[red edge] (B3LLRdot) -- ($(L4lab)+(0,0.45)$);

\draw[red edge] (B3LRLdot) -- ($(L6lab)+(0,0.45)$);
\draw[red edge] (B3LRRdot) -- ($(L8lab)+(0,0.45)$);

\draw[red edge] (B3RLLdot) -- ($(L10lab)+(0,0.45)$);
\draw[red edge] (B3RLRdot) -- ($(L11lab)+(0,0.45)$);

\draw[red edge] (B3RRLdot) -- ($(L13lab)+(0,0.45)$);
\draw[red edge] (B3RRRdot) -- ($(L15lab)+(0,0.45)$);

\node[minlabel] at ($(A2LL)!0.5!(B3LLL)+(-0.65,0.22)$) {$\min\ 2$};
\node[minlabel] at ($(A2LL)!0.5!(B3LLR)+(0.65,0.22)$)  {$\min\ 7$};

\node[minlabel] at ($(A2LR)!0.5!(B3LRL)+(-0.65,0.22)$) {$\min\ 1$};
\node[minlabel] at ($(A2LR)!0.5!(B3LRR)+(0.65,0.22)$)  {$\min\ 4$};

\node[minlabel] at ($(A2RL)!0.5!(B3RLL)+(-0.65,0.22)$) {$\min\ 0$};
\node[minlabel] at ($(A2RL)!0.5!(B3RLR)+(0.65,0.22)$)  {$\min\ 8$};

\node[minlabel] at ($(A2RR)!0.5!(B3RRL)+(-0.65,0.22)$) {$\min\ 3$};
\node[minlabel] at ($(A2RR)!0.5!(B3RRR)+(0.65,0.22)$)  {$\min\ 6$};

\draw[green ul] ($(L3lab)+(-0.38,-0.58)$) -- ($(L4lab)+(0.38,-0.58)$);
\draw[green ul] ($(L7lab)+(-0.38,-0.58)$) -- ($(L8lab)+(0.38,-0.58)$);
\draw[green ul] ($(L11lab)+(-0.38,-0.58)$) -- ($(L12lab)+(0.38,-0.58)$);
\draw[green ul] ($(L15lab)+(-0.38,-0.58)$) -- ($(L16lab)+(0.38,-0.58)$);

\draw[red ul] ($(L4lab)+(-0.3,-0.43)$) -- ($(L4lab)+(0.3,-0.43)$);
\draw[red ul] ($(L8lab)+(-0.3,-0.43)$) -- ($(L8lab)+(0.3,-0.43)$);
\draw[red ul] ($(L11lab)+(-0.3,-0.43)$) -- ($(L11lab)+(0.3,-0.43)$);
\draw[red ul] ($(L15lab)+(-0.3,-0.43)$) -- ($(L15lab)+(0.3,-0.43)$);

\end{tikzpicture}
}
\caption{Level-2 recursive reasoning in the two-level-per-player game. Red arrows indicate the realized choices at the $A$-nodes and $B$-nodes under optimal play. The green $\min$ labels show the guaranteed payoff of each action at a level-2 $A$-node. Dark red underlines mark the outcome that is realized under optimal play from that node, while dark green underlines mark the outcomes that remain possible for player $A$ along the chosen direction.}
\end{figure}
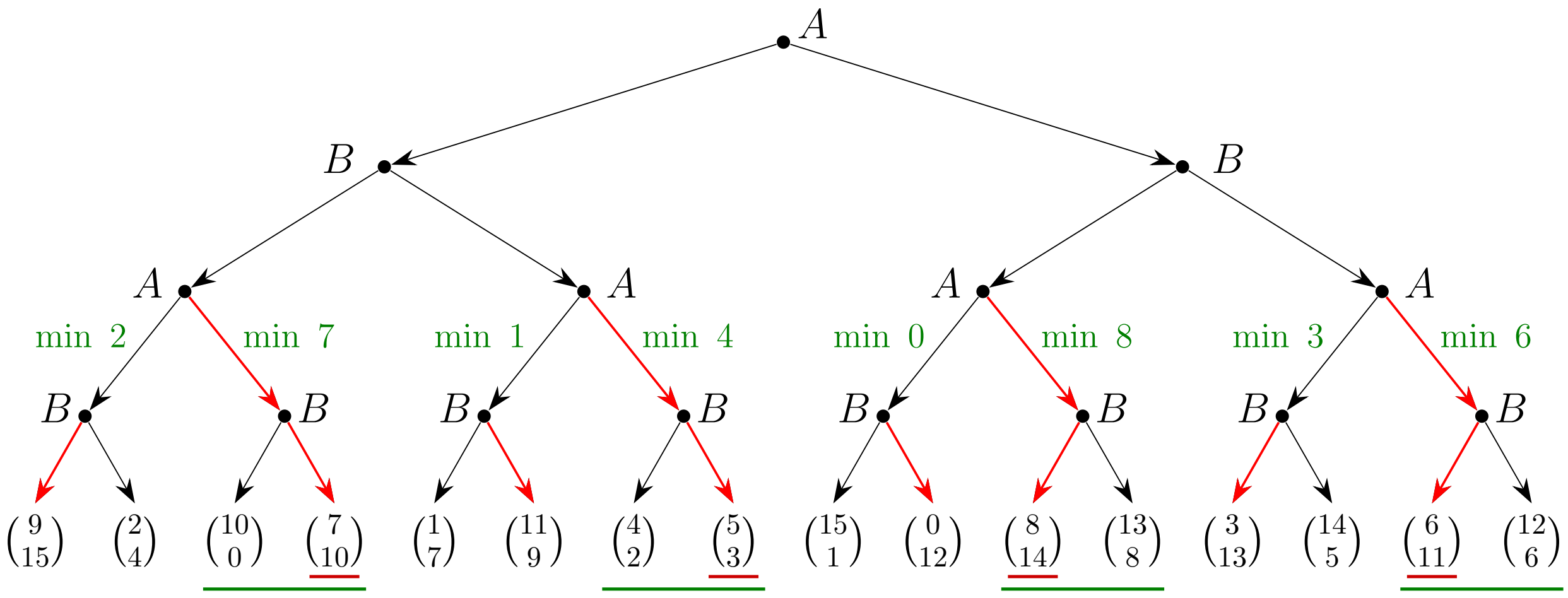

\medskip
\noindent
\textbf{Level-1 reasoning.}
At a level-1 $B$-node, $B$ selects the direction with the higher realized payoff. From player $A$'s perspective, however, the outcomes in both subtrees remain possible unless ruled out by a revelation that eliminates the outcomes in one of the subtrees from $A$'s consideration. At the root, player $A$'s guaranteed payoff for each direction is the minimum of $A$'s payoffs over the outcomes that remain possible at the corresponding level-1 $B-$node. Thus, $A$ minimizes not over all terminal outcomes, but only over those that remain possible after carrying out recursive reasoning.

Figure 6 extends the level-2 annotations from Figure 5 to the full tree by propagating these computations upward.

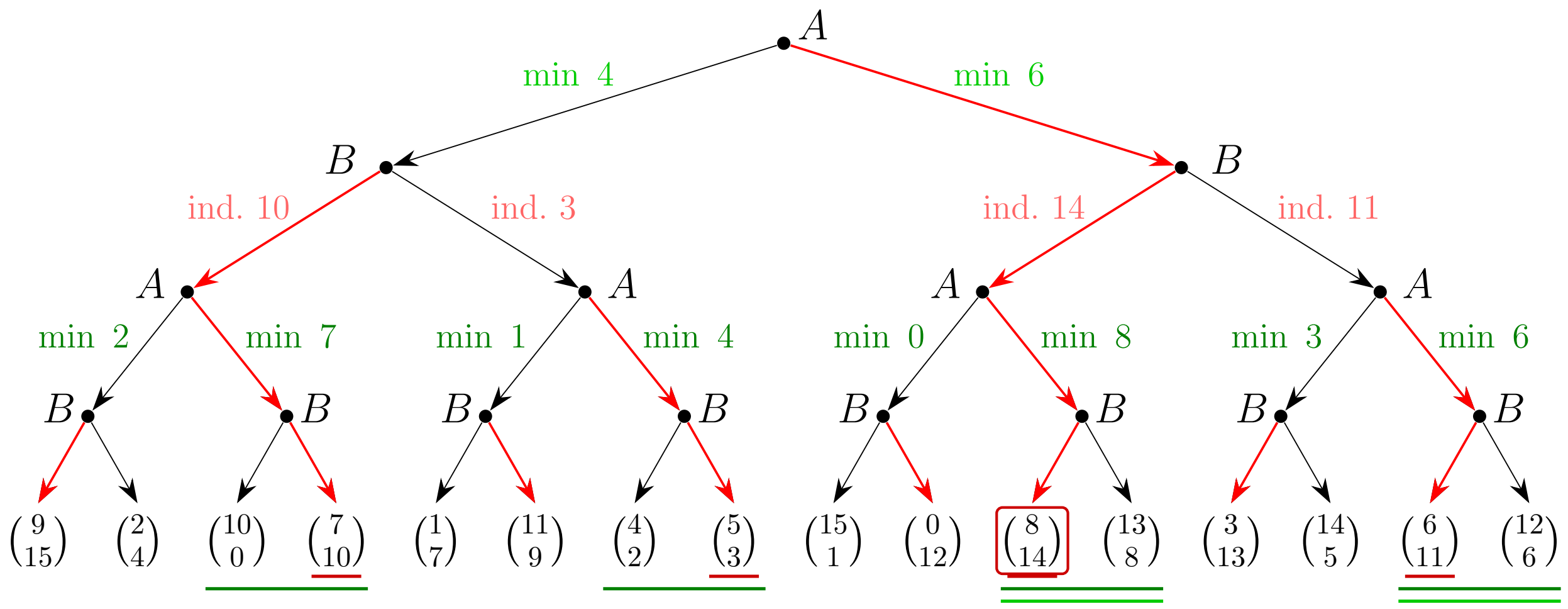
\begin{figure}[H]
\centering
\resizebox{\textwidth}{!}{%
\begin{tikzpicture}[
    x=1cm, y=1cm,
    >={Stealth[length=3mm,width=2mm]},
    every node/.style={inner sep=0pt},
    player/.style={font=\Large},
    payoff/.style={font=\Large},
    minlabel/.style={font=\large\color{green!50!black}},
    rootminlabel/.style={font=\large\color{green!80!black}},
    indlabel/.style={font=\large\color{red!60}},
    node dot/.style={circle, fill=black, inner sep=1.6pt},
    black edge/.style={->, line width=0.4pt, black},
    red edge/.style={->, line width=0.8pt, red},
    green ul/.style={line width=1.1pt, color=green!50!black},
    red ul/.style={line width=1.1pt, color=red!80!black},
    lightgreen ul/.style={line width=1.1pt, color=green!80!black},
    lightred ul/.style={line width=1.1pt, color=red!45!black}
]

\coordinate (A0) at (0,6.0);

\coordinate (B1L) at (-4.8,4.5);
\coordinate (B1R) at ( 4.8,4.5);

\coordinate (A2LL) at (-7.2,3.0);
\coordinate (A2LR) at (-2.4,3.0);
\coordinate (A2RL) at ( 2.4,3.0);
\coordinate (A2RR) at ( 7.2,3.0);

\coordinate (B3LLL) at (-8.4,1.5);
\coordinate (B3LLR) at (-6.0,1.5);

\coordinate (B3LRL) at (-3.6,1.5);
\coordinate (B3LRR) at (-1.2,1.5);

\coordinate (B3RLL) at ( 1.2,1.5);
\coordinate (B3RLR) at ( 3.6,1.5);

\coordinate (B3RRL) at ( 6.0,1.5);
\coordinate (B3RRR) at ( 8.4,1.5);

\coordinate (L1)  at (-9.0,0.0);
\coordinate (L2)  at (-7.8,0.0);
\coordinate (L3)  at (-6.6,0.0);
\coordinate (L4)  at (-5.4,0.0);

\coordinate (L5)  at (-4.2,0.0);
\coordinate (L6)  at (-3.0,0.0);
\coordinate (L7)  at (-1.8,0.0);
\coordinate (L8)  at (-0.6,0.0);

\coordinate (L9)  at ( 0.6,0.0);
\coordinate (L10) at ( 1.8,0.0);
\coordinate (L11) at ( 3.0,0.0);
\coordinate (L12) at ( 4.2,0.0);

\coordinate (L13) at ( 5.4,0.0);
\coordinate (L14) at ( 6.6,0.0);
\coordinate (L15) at ( 7.8,0.0);
\coordinate (L16) at ( 9.0,0.0);

\node[node dot] (A0dot)    at (A0)    {};
\node[node dot] (B1Ldot)   at (B1L)   {};
\node[node dot] (B1Rdot)   at (B1R)   {};

\node[node dot] (A2LLdot)  at (A2LL)  {};
\node[node dot] (A2LRdot)  at (A2LR)  {};
\node[node dot] (A2RLdot)  at (A2RL)  {};
\node[node dot] (A2RRdot)  at (A2RR)  {};

\node[node dot] (B3LLLdot) at (B3LLL) {};
\node[node dot] (B3LLRdot) at (B3LLR) {};
\node[node dot] (B3LRLdot) at (B3LRL) {};
\node[node dot] (B3LRRdot) at (B3LRR) {};
\node[node dot] (B3RLLdot) at (B3RLL) {};
\node[node dot] (B3RLRdot) at (B3RLR) {};
\node[node dot] (B3RRLdot) at (B3RRL) {};
\node[node dot] (B3RRRdot) at (B3RRR) {};

\node[player] at ($(A0)+(0.35,0.22)$) {$A$};

\node[player] at ($(B1L)+(-0.55,0.10)$) {$B$};
\node[player] at ($(B1R)+(0.55,0.10)$)  {$B$};

\node[player] at ($(A2LL)+(-0.45,0.10)$) {$A$};
\node[player] at ($(A2LR)+(0.45,0.10)$)  {$A$};
\node[player] at ($(A2RL)+(-0.45,0.10)$) {$A$};
\node[player] at ($(A2RR)+(0.45,0.10)$)  {$A$};

\node[player] at ($(B3LLL)+(-0.35,0.10)$) {$B$};
\node[player] at ($(B3LLR)+(0.35,0.10)$)  {$B$};
\node[player] at ($(B3LRL)+(-0.35,0.10)$) {$B$};
\node[player] at ($(B3LRR)+(0.35,0.10)$)  {$B$};
\node[player] at ($(B3RLL)+(-0.35,0.10)$) {$B$};
\node[player] at ($(B3RLR)+(0.35,0.10)$)  {$B$};
\node[player] at ($(B3RRL)+(-0.35,0.10)$) {$B$};
\node[player] at ($(B3RRR)+(0.35,0.10)$)  {$B$};

\node[payoff] (L1lab)  at (L1)  {$\binom{9}{15}$};
\node[payoff] (L2lab)  at (L2)  {$\binom{2}{4}$};
\node[payoff] (L3lab)  at (L3)  {$\binom{10}{0}$};
\node[payoff] (L4lab)  at (L4)  {$\binom{7}{10}$};

\node[payoff] (L5lab)  at (L5)  {$\binom{1}{7}$};
\node[payoff] (L6lab)  at (L6)  {$\binom{11}{9}$};
\node[payoff] (L7lab)  at (L7)  {$\binom{4}{2}$};
\node[payoff] (L8lab)  at (L8)  {$\binom{5}{3}$};

\node[payoff] (L9lab)  at (L9)  {$\binom{15}{1}$};
\node[payoff] (L10lab) at (L10) {$\binom{0}{12}$};
\node[payoff] (L11lab) at (L11) {$\binom{8}{14}$};
\node[payoff] (L12lab) at (L12) {$\binom{13}{8}$};

\node[payoff] (L13lab) at (L13) {$\binom{3}{13}$};
\node[payoff] (L14lab) at (L14) {$\binom{14}{5}$};
\node[payoff] (L15lab) at (L15) {$\binom{6}{11}$};
\node[payoff] (L16lab) at (L16) {$\binom{12}{6}$};

\draw[black edge] (A0dot) -- (B1Ldot);
\draw[red edge] (A0dot) -- (B1Rdot);

\draw[red edge] (B1Ldot) -- (A2LLdot);
\draw[black edge] (B1Ldot) -- (A2LRdot);
\draw[red edge] (B1Rdot) -- (A2RLdot);
\draw[black edge] (B1Rdot) -- (A2RRdot);

\draw[black edge] (A2LLdot) -- (B3LLLdot);
\draw[black edge] (A2LLdot) -- (B3LLRdot);
\draw[black edge] (A2LRdot) -- (B3LRLdot);
\draw[black edge] (A2LRdot) -- (B3LRRdot);
\draw[black edge] (A2RLdot) -- (B3RLLdot);
\draw[black edge] (A2RLdot) -- (B3RLRdot);
\draw[black edge] (A2RRdot) -- (B3RRLdot);
\draw[black edge] (A2RRdot) -- (B3RRRdot);

\draw[black edge] (B3LLLdot) -- ($(L1lab)+(0,0.45)$);
\draw[black edge] (B3LLLdot) -- ($(L2lab)+(0,0.45)$);
\draw[black edge] (B3LLRdot) -- ($(L3lab)+(0,0.45)$);
\draw[black edge] (B3LLRdot) -- ($(L4lab)+(0,0.45)$);

\draw[black edge] (B3LRLdot) -- ($(L5lab)+(0,0.45)$);
\draw[black edge] (B3LRLdot) -- ($(L6lab)+(0,0.45)$);
\draw[black edge] (B3LRRdot) -- ($(L7lab)+(0,0.45)$);
\draw[black edge] (B3LRRdot) -- ($(L8lab)+(0,0.45)$);

\draw[black edge] (B3RLLdot) -- ($(L9lab)+(0,0.45)$);
\draw[black edge] (B3RLLdot) -- ($(L10lab)+(0,0.45)$);
\draw[black edge] (B3RLRdot) -- ($(L11lab)+(0,0.45)$);
\draw[black edge] (B3RLRdot) -- ($(L12lab)+(0,0.45)$);

\draw[black edge] (B3RRLdot) -- ($(L13lab)+(0,0.45)$);
\draw[black edge] (B3RRLdot) -- ($(L14lab)+(0,0.45)$);
\draw[black edge] (B3RRRdot) -- ($(L15lab)+(0,0.45)$);
\draw[black edge] (B3RRRdot) -- ($(L16lab)+(0,0.45)$);

\draw[red edge] (A2LLdot) -- (B3LLRdot);
\draw[red edge] (A2LRdot) -- (B3LRRdot);
\draw[red edge] (A2RLdot) -- (B3RLRdot);
\draw[red edge] (A2RRdot) -- (B3RRRdot);

\draw[red edge] (B3LLLdot) -- ($(L1lab)+(0,0.45)$);
\draw[red edge] (B3LLRdot) -- ($(L4lab)+(0,0.45)$);

\draw[red edge] (B3LRLdot) -- ($(L6lab)+(0,0.45)$);
\draw[red edge] (B3LRRdot) -- ($(L8lab)+(0,0.45)$);

\draw[red edge] (B3RLLdot) -- ($(L10lab)+(0,0.45)$);
\draw[red edge] (B3RLRdot) -- ($(L11lab)+(0,0.45)$);

\draw[red edge] (B3RRLdot) -- ($(L13lab)+(0,0.45)$);
\draw[red edge] (B3RRRdot) -- ($(L15lab)+(0,0.45)$);

\node[indlabel] at ($(B1L)!0.5!(A2LL)+(-0.58,0.28)$) {$\mathrm{ind.}\ 10$};
\node[indlabel] at ($(B1L)!0.5!(A2LR)+(0.58,0.28)$)  {$\mathrm{ind.}\ 3$};

\node[indlabel] at ($(B1R)!0.5!(A2RL)+(-0.58,0.28)$) {$\mathrm{ind.}\ 14$};
\node[indlabel] at ($(B1R)!0.5!(A2RR)+(0.58,0.28)$)  {$\mathrm{ind.}\ 11$};

\node[rootminlabel] at ($(A0)!0.5!(B1L)+(-0.20,0.38)$) {$\min\ 4$};
\node[rootminlabel] at ($(A0)!0.5!(B1R)+(0.20,0.38)$)  {$\min\ 6$};

\node[minlabel] at ($(A2LL)!0.5!(B3LLL)+(-0.65,0.22)$) {$\min\ 2$};
\node[minlabel] at ($(A2LL)!0.5!(B3LLR)+(0.65,0.22)$)  {$\min\ 7$};

\node[minlabel] at ($(A2LR)!0.5!(B3LRL)+(-0.65,0.22)$) {$\min\ 1$};
\node[minlabel] at ($(A2LR)!0.5!(B3LRR)+(0.65,0.22)$)  {$\min\ 4$};

\node[minlabel] at ($(A2RL)!0.5!(B3RLL)+(-0.65,0.22)$) {$\min\ 0$};
\node[minlabel] at ($(A2RL)!0.5!(B3RLR)+(0.65,0.22)$)  {$\min\ 8$};

\node[minlabel] at ($(A2RR)!0.5!(B3RRL)+(-0.65,0.22)$) {$\min\ 3$};
\node[minlabel] at ($(A2RR)!0.5!(B3RRR)+(0.65,0.22)$)  {$\min\ 6$};

\draw[green ul] ($(L3lab)+(-0.38,-0.58)$) -- ($(L4lab)+(0.38,-0.58)$);
\draw[green ul] ($(L7lab)+(-0.38,-0.58)$) -- ($(L8lab)+(0.38,-0.58)$);
\draw[green ul] ($(L11lab)+(-0.38,-0.58)$) -- ($(L12lab)+(0.38,-0.58)$);
\draw[green ul] ($(L15lab)+(-0.38,-0.58)$) -- ($(L16lab)+(0.38,-0.58)$);

\draw[line width=1.1pt, color=green!80!black]
    ($(L11lab)+(-0.38,-0.73)$) -- ($(L12lab)+(0.38,-0.73)$);
\draw[line width=1.1pt, color=green!80!black]
    ($(L15lab)+(-0.38,-0.73)$) -- ($(L16lab)+(0.38,-0.73)$);

\draw[red ul] ($(L4lab)+(-0.3,-0.43)$) -- ($(L4lab)+(0.3,-0.43)$);
\draw[red ul] ($(L8lab)+(-0.3,-0.43)$) -- ($(L8lab)+(0.3,-0.43)$);
\draw[red ul] ($(L11lab)+(-0.3,-0.43)$) -- ($(L11lab)+(0.3,-0.43)$);
\draw[red ul] ($(L15lab)+(-0.3,-0.43)$) -- ($(L15lab)+(0.3,-0.43)$);

\draw[draw=red!80!black, line width=0.9pt, rounded corners=2pt]
    ($(L11lab)+(-0.43,0.40)$) rectangle ($(L11lab)+(0.43,-0.40)$);

\end{tikzpicture}
}
\caption{Full recursive reasoning in the two-level-per-player game. The dark green $\min$ labels, dark green underlines, and dark red underlines are the same as in Figure~5, recording the guaranteed payoffs, the outcomes that remain possible for player $A$, and the realized outcomes at the level-2 $A$-nodes. The light red $\mathrm{ind.}$ labels at the level-1 $B$-nodes record the realized outcomes propagated upward from the two level-2 subtrees. The light green $\min$ labels at the root show the resulting guaranteed payoffs for player $A$ at the root. The light green underlines mark the outcomes that remain possible for player $A$ at the root, and the red rounded rectangle highlights the outcome ultimately realized under optimal play from the root.}
\end{figure}

To conclude, we illustrate how truthful revelations by $B$ alter the recursive reasoning process and improve $B$'s payoff. Figure 7 shows how two truthful revelations along the path to $(9,15)$ induce this outcome. Each revelation prunes outcomes inconsistent with the committed move: at level 2, an individual outcome is removed, while at level 1 the pruning propagates upward and eliminates an entire $A$-rooted subtree from consideration. The resulting reduction in feasible continuations increases $A$'s guaranteed payoff along the left branches from the $A$-nodes along the path, so $A$ switches and the play follows the path to $(9,15)$.


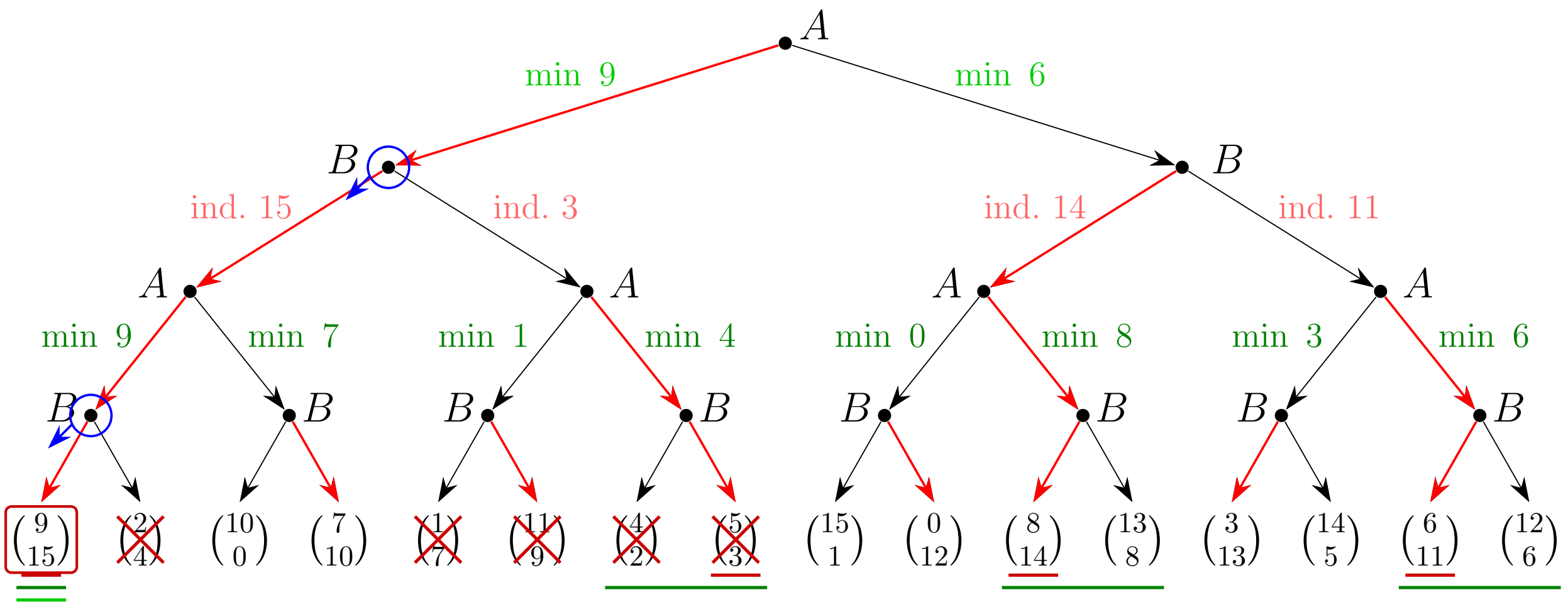
\begin{figure}[h]
\centering
\resizebox{\textwidth}{!}{%
\begin{tikzpicture}[
    x=1cm, y=1cm,
    >={Stealth[length=3mm,width=2mm]},
    every node/.style={inner sep=0pt},
    red x/.style={red!80!black, line width=1.1pt},
    player/.style={font=\Large},
    payoff/.style={font=\Large},
    minlabel/.style={font=\large\color{green!50!black}},
    rootminlabel/.style={font=\large\color{green!80!black}},
    indlabel/.style={font=\large\color{red!60}},
    node dot/.style={circle, fill=black, inner sep=1.6pt},
    black edge/.style={->, line width=0.4pt, black},
    red edge/.style={->, line width=0.8pt, red},
    green ul/.style={line width=1.1pt, color=green!50!black},
    red ul/.style={line width=1.1pt, color=red!80!black},
    lightgreen ul/.style={line width=1.1pt, color=green!80!black}
]

\coordinate (A0) at (0,6.0);

\coordinate (B1L) at (-4.8,4.5);
\coordinate (B1R) at ( 4.8,4.5);

\coordinate (A2LL) at (-7.2,3.0);
\coordinate (A2LR) at (-2.4,3.0);
\coordinate (A2RL) at ( 2.4,3.0);
\coordinate (A2RR) at ( 7.2,3.0);

\coordinate (B3LLL) at (-8.4,1.5);
\coordinate (B3LLR) at (-6.0,1.5);

\coordinate (B3LRL) at (-3.6,1.5);
\coordinate (B3LRR) at (-1.2,1.5);

\coordinate (B3RLL) at ( 1.2,1.5);
\coordinate (B3RLR) at ( 3.6,1.5);

\coordinate (B3RRL) at ( 6.0,1.5);
\coordinate (B3RRR) at ( 8.4,1.5);

\coordinate (L1)  at (-9.0,0.0);
\coordinate (L2)  at (-7.8,0.0);
\coordinate (L3)  at (-6.6,0.0);
\coordinate (L4)  at (-5.4,0.0);

\coordinate (L5)  at (-4.2,0.0);
\coordinate (L6)  at (-3.0,0.0);
\coordinate (L7)  at (-1.8,0.0);
\coordinate (L8)  at (-0.6,0.0);

\coordinate (L9)  at ( 0.6,0.0);
\coordinate (L10) at ( 1.8,0.0);
\coordinate (L11) at ( 3.0,0.0);
\coordinate (L12) at ( 4.2,0.0);

\coordinate (L13) at ( 5.4,0.0);
\coordinate (L14) at ( 6.6,0.0);
\coordinate (L15) at ( 7.8,0.0);
\coordinate (L16) at ( 9.0,0.0);

\node[node dot] (A0dot)    at (A0)    {};
\node[node dot] (B1Ldot)   at (B1L)   {};
\node[node dot] (B1Rdot)   at (B1R)   {};

\node[node dot] (A2LLdot)  at (A2LL)  {};
\node[node dot] (A2LRdot)  at (A2LR)  {};
\node[node dot] (A2RLdot)  at (A2RL)  {};
\node[node dot] (A2RRdot)  at (A2RR)  {};

\node[node dot] (B3LLLdot) at (B3LLL) {};
\node[node dot] (B3LLRdot) at (B3LLR) {};
\node[node dot] (B3LRLdot) at (B3LRL) {};
\node[node dot] (B3LRRdot) at (B3LRR) {};
\node[node dot] (B3RLLdot) at (B3RLL) {};
\node[node dot] (B3RLRdot) at (B3RLR) {};
\node[node dot] (B3RRLdot) at (B3RRL) {};
\node[node dot] (B3RRRdot) at (B3RRR) {};

\node[player] at ($(A0)+(0.35,0.22)$) {$A$};

\node[player] at ($(B1L)+(-0.55,0.10)$) {$B$};
\node[player] at ($(B1R)+(0.55,0.10)$)  {$B$};

\node[player] at ($(A2LL)+(-0.45,0.10)$) {$A$};
\node[player] at ($(A2LR)+(0.45,0.10)$)  {$A$};
\node[player] at ($(A2RL)+(-0.45,0.10)$) {$A$};
\node[player] at ($(A2RR)+(0.45,0.10)$)  {$A$};

\node[player] at ($(B3LLL)+(-0.35,0.10)$) {$B$};
\node[player] at ($(B3LLR)+(0.35,0.10)$)  {$B$};
\node[player] at ($(B3LRL)+(-0.35,0.10)$) {$B$};
\node[player] at ($(B3LRR)+(0.35,0.10)$)  {$B$};
\node[player] at ($(B3RLL)+(-0.35,0.10)$) {$B$};
\node[player] at ($(B3RLR)+(0.35,0.10)$)  {$B$};
\node[player] at ($(B3RRL)+(-0.35,0.10)$) {$B$};
\node[player] at ($(B3RRR)+(0.35,0.10)$)  {$B$};

\node[payoff] (L1lab)  at (L1)  {$\binom{9}{15}$};
\node[payoff] (L2lab)  at (L2)  {$\binom{2}{4}$};
\node[payoff] (L3lab)  at (L3)  {$\binom{10}{0}$};
\node[payoff] (L4lab)  at (L4)  {$\binom{7}{10}$};

\node[payoff] (L5lab)  at (L5)  {$\binom{1}{7}$};
\node[payoff] (L6lab)  at (L6)  {$\binom{11}{9}$};
\node[payoff] (L7lab)  at (L7)  {$\binom{4}{2}$};
\node[payoff] (L8lab)  at (L8)  {$\binom{5}{3}$};

\node[payoff] (L9lab)  at (L9)  {$\binom{15}{1}$};
\node[payoff] (L10lab) at (L10) {$\binom{0}{12}$};
\node[payoff] (L11lab) at (L11) {$\binom{8}{14}$};
\node[payoff] (L12lab) at (L12) {$\binom{13}{8}$};

\node[payoff] (L13lab) at (L13) {$\binom{3}{13}$};
\node[payoff] (L14lab) at (L14) {$\binom{14}{5}$};
\node[payoff] (L15lab) at (L15) {$\binom{6}{11}$};
\node[payoff] (L16lab) at (L16) {$\binom{12}{6}$};

\draw[black edge] (A0dot) -- (B1Ldot);
\draw[black edge] (A0dot) -- (B1Rdot);

\draw[black edge] (B1Ldot) -- (A2LLdot);
\draw[black edge] (B1Ldot) -- (A2LRdot);
\draw[black edge] (B1Rdot) -- (A2RLdot);
\draw[black edge] (B1Rdot) -- (A2RRdot);

\draw[black edge] (A2LLdot) -- (B3LLLdot);
\draw[black edge] (A2LLdot) -- (B3LLRdot);
\draw[black edge] (A2LRdot) -- (B3LRLdot);
\draw[black edge] (A2LRdot) -- (B3LRRdot);
\draw[black edge] (A2RLdot) -- (B3RLLdot);
\draw[black edge] (A2RLdot) -- (B3RLRdot);
\draw[black edge] (A2RRdot) -- (B3RRLdot);
\draw[black edge] (A2RRdot) -- (B3RRRdot);

\draw[black edge] (B3LLLdot) -- ($(L1lab)+(0,0.45)$);
\draw[black edge] (B3LLLdot) -- ($(L2lab)+(0,0.45)$);
\draw[black edge] (B3LLRdot) -- ($(L3lab)+(0,0.45)$);
\draw[black edge] (B3LLRdot) -- ($(L4lab)+(0,0.45)$);

\draw[black edge] (B3LRLdot) -- ($(L5lab)+(0,0.45)$);
\draw[black edge] (B3LRLdot) -- ($(L6lab)+(0,0.45)$);
\draw[black edge] (B3LRRdot) -- ($(L7lab)+(0,0.45)$);
\draw[black edge] (B3LRRdot) -- ($(L8lab)+(0,0.45)$);

\draw[black edge] (B3RLLdot) -- ($(L9lab)+(0,0.45)$);
\draw[black edge] (B3RLLdot) -- ($(L10lab)+(0,0.45)$);
\draw[black edge] (B3RLRdot) -- ($(L11lab)+(0,0.45)$);
\draw[black edge] (B3RLRdot) -- ($(L12lab)+(0,0.45)$);

\draw[black edge] (B3RRLdot) -- ($(L13lab)+(0,0.45)$);
\draw[black edge] (B3RRLdot) -- ($(L14lab)+(0,0.45)$);
\draw[black edge] (B3RRRdot) -- ($(L15lab)+(0,0.45)$);
\draw[black edge] (B3RRRdot) -- ($(L16lab)+(0,0.45)$);

\draw[red edge] (A2LLdot) -- (B3LLLdot);
\draw[red edge] (A2LRdot) -- (B3LRRdot);
\draw[red edge] (A2RLdot) -- (B3RLRdot);
\draw[red edge] (A2RRdot) -- (B3RRRdot);

\draw[red edge] (B3LLLdot) -- ($(L1lab)+(0,0.45)$);
\draw[red edge] (B3LLRdot) -- ($(L4lab)+(0,0.45)$);

\draw[red edge] (B3LRLdot) -- ($(L6lab)+(0,0.45)$);
\draw[red edge] (B3LRRdot) -- ($(L8lab)+(0,0.45)$);

\draw[red edge] (B3RLLdot) -- ($(L10lab)+(0,0.45)$);
\draw[red edge] (B3RLRdot) -- ($(L11lab)+(0,0.45)$);

\draw[red edge] (B3RRLdot) -- ($(L13lab)+(0,0.45)$);
\draw[red edge] (B3RRRdot) -- ($(L15lab)+(0,0.45)$);

\draw[red edge] (B1Ldot) -- (A2LLdot);
\draw[red edge] (B1Rdot) -- (A2RLdot);
\draw[red edge] (A0dot) -- (B1Ldot);

\drawrevcircle{B1Ldot}
\drawrevleft{B1Ldot}

\drawrevcircle{B3LLLdot}
\drawrevleft{B3LLLdot}

\node[indlabel] at ($(B1L)!0.5!(A2LL)+(-0.58,0.28)$) {$\mathrm{ind.}\ 15$};
\node[indlabel] at ($(B1L)!0.5!(A2LR)+(0.58,0.28)$)  {$\mathrm{ind.}\ 3$};

\node[indlabel] at ($(B1R)!0.5!(A2RL)+(-0.58,0.28)$) {$\mathrm{ind.}\ 14$};
\node[indlabel] at ($(B1R)!0.5!(A2RR)+(0.58,0.28)$)  {$\mathrm{ind.}\ 11$};

\node[rootminlabel] at ($(A0)!0.5!(B1L)+(-0.20,0.38)$) {$\min\ 9$};
\node[rootminlabel] at ($(A0)!0.5!(B1R)+(0.20,0.38)$)  {$\min\ 6$};

\node[minlabel] at ($(A2LL)!0.5!(B3LLL)+(-0.65,0.22)$) {$\min\ 9$};
\node[minlabel] at ($(A2LL)!0.5!(B3LLR)+(0.65,0.22)$)  {$\min\ 7$};

\node[minlabel] at ($(A2LR)!0.5!(B3LRL)+(-0.65,0.22)$) {$\min\ 1$};
\node[minlabel] at ($(A2LR)!0.5!(B3LRR)+(0.65,0.22)$)  {$\min\ 4$};

\node[minlabel] at ($(A2RL)!0.5!(B3RLL)+(-0.65,0.22)$) {$\min\ 0$};
\node[minlabel] at ($(A2RL)!0.5!(B3RLR)+(0.65,0.22)$)  {$\min\ 8$};

\node[minlabel] at ($(A2RR)!0.5!(B3RRL)+(-0.65,0.22)$) {$\min\ 3$};
\node[minlabel] at ($(A2RR)!0.5!(B3RRR)+(0.65,0.22)$)  {$\min\ 6$};

\draw[green ul] ($(L1lab)+(-0.30,-0.58)$) -- ($(L1lab)+(0.30,-0.58)$);
\draw[green ul] ($(L7lab)+(-0.38,-0.58)$) -- ($(L8lab)+(0.38,-0.58)$);
\draw[green ul] ($(L11lab)+(-0.38,-0.58)$) -- ($(L12lab)+(0.38,-0.58)$);
\draw[green ul] ($(L15lab)+(-0.38,-0.58)$) -- ($(L16lab)+(0.38,-0.58)$);

\draw[lightgreen ul] ($(L1lab)+(-0.30,-0.73)$) -- ($(L1lab)+(0.30,-0.73)$);

\draw[red ul] ($(L1lab)+(-0.24,-0.43)$) -- ($(L1lab)+(0.24,-0.43)$);
\draw[red ul] ($(L8lab)+(-0.30,-0.43)$) -- ($(L8lab)+(0.30,-0.43)$);
\draw[red ul] ($(L11lab)+(-0.30,-0.43)$) -- ($(L11lab)+(0.30,-0.43)$);
\draw[red ul] ($(L15lab)+(-0.30,-0.43)$) -- ($(L15lab)+(0.30,-0.43)$);

\draw[draw=red!80!black, line width=0.9pt, rounded corners=2pt]
    ($(L1lab)+(-0.43,0.40)$) rectangle ($(L1lab)+(0.43,-0.40)$);


\draw[red x] ($(L2lab.center)+(-0.28,-0.28)$) -- ($(L2lab.center)+(0.28,0.28)$);
\draw[red x] ($(L2lab.center)+(-0.28,0.28)$) -- ($(L2lab.center)+(0.28,-0.28)$);

\draw[red x] ($(L5lab.center)+(-0.28,-0.28)$) -- ($(L5lab.center)+(0.28,0.28)$);
\draw[red x] ($(L5lab.center)+(-0.28,0.28)$) -- ($(L5lab.center)+(0.28,-0.28)$);

\draw[red x] ($(L6lab.center)+(-0.28,-0.28)$) -- ($(L6lab.center)+(0.28,0.28)$);
\draw[red x] ($(L6lab.center)+(-0.28,0.28)$) -- ($(L6lab.center)+(0.28,-0.28)$);

\draw[red x] ($(L7lab.center)+(-0.28,-0.28)$) -- ($(L7lab.center)+(0.28,0.28)$);
\draw[red x] ($(L7lab.center)+(-0.28,0.28)$) -- ($(L7lab.center)+(0.28,-0.28)$);

\draw[red x] ($(L8lab.center)+(-0.28,-0.28)$) -- ($(L8lab.center)+(0.28,0.28)$);
\draw[red x] ($(L8lab.center)+(-0.28,0.28)$) -- ($(L8lab.center)+(0.28,-0.28)$);

\end{tikzpicture}
}
\caption{Full recursive reasoning in the two-level-per-player game after revealing along the path to $(9,15)$. The blue circles and arrows indicate player $B$'s truthful commitments at the two revealed $B$-nodes. The red X marks show the outcomes pruned from player $A$'s consideration: the lower revelation removes the outcome $(2,4)$, while the higher revelation removes the $A$-rooted subtree leading to $(1,7)$, $(11,9)$, $(4,2)$, and $(5,3)$. These prunings change the recursive computation of guaranteed payoffs: the dark green and dark red annotations play the same role as in Figure~6, updated to reflect the new computation. The light red $\mathrm{ind.}$ labels record the realized outcomes propagated upward from the level-2 subtrees, the light green $\min$ labels at the root show the resulting guaranteed payoffs for player $A$, and the light green underline marks the outcomes that remain possible for player $A$ at the root. The resulting guarantee on the left branch is $9$, exceeding the right-branch guarantee of $6$, so player $A$ moves left and the red rounded rectangle marks the realized outcome $(9,15)$.}
\end{figure}

\section{Main Result: Epistemic Argument}

\subsection{From Examples to Paths}

As seen in Section 3, player $A$ evaluates actions using guaranteed payoffs, while player $B$ can reveal truthful commitments that eliminate outcomes inconsistent with those commitments. Such revelations influence behavior not by directly selecting actions, but by pruning the set of continuations that player $A$ considers possible.

The examples in Section 3 show that, in the cases we construct, successful revelations occur along a path leading to the outcome that is ultimately realized. This suggests that the key object is not the set of revelations itself, but rather the path of play that can be sustained under the feasible outcomes induced by those revelations. As anticipated in the introduction, our analysis proceeds by characterizing which paths remain optimal for player $A$ under the induced feasible outcome sets.

For a path to be sustained, it must satisfy a local optimality condition: when player $B$ reveals the moves along the path, then at every $A$-node along the path, the guaranteed payoff from following the path is strictly higher than the guaranteed payoff from any deviation at that node. While stated here under maximin reasoning, this condition depends only on how feasible outcome sets determine guarantees, and thus reflects a more general structural phenomenon.

\subsection{Definitions and Main Result}

We formalize the above idea as follows.

\begin{definition}[Possible path]
A path $P$ is \emph{possible} if, when player $B$ reveals the moves along $P$, then at every $A$-node along $P$, the guaranteed payoff from following $P$ is strictly higher than the guaranteed payoff from any deviation at that node.\end{definition}

Thus, a possible path is one that is consistent with player $A$'s decision rule under the assumption that the moves along the path are known. In particular, this definition depends only on player $A$'s reasoning.

\begin{definition}[Inducible outcome]
An outcome $X$ is \emph{inducible} if it is the terminal outcome of some possible path $P$.
\end{definition}

Thus, player $B$'s problem reduces to selecting a possible path and revealing the corresponding moves to player $A$. By truthfulness, these revelations eliminate outcomes inconsistent with the path, ensuring that player $A$'s reasoning leads to the prescribed outcome.

This yields the main result of the paper: every path satisfying this local condition can be implemented via truthful revelation, with a bounded number of disclosures.

\begin{theorem}
Every possible path $P$ can be implemented by player $B$ through truthful revelation. If $B$ reveals all of their moves along $P$ to player $A$, then the resulting play of the game follows $P$. Moreover, this requires at most $n$ revelations.
\end{theorem}

\begin{proof}
Fix such a possible path \(P\). By Definition 1, if player \(B\)'s moves along \(P\) are revealed to player \(A\), then at every \(A\)-node along \(P\), following \(P\) yields a strictly higher guaranteed payoff than any deviation at that node.

Player \(B\) reveals their moves along \(P\). Since the revelations are truthful, each such revelation removes from \(A\)'s consideration the outcomes inconsistent with the revealed move at that \(B\)-node. Other outcomes not ruled out by these revealed moves may remain possible.

Under the resulting set of possible outcomes, Definition 1 implies that player \(A\)'s prescribed move along \(P\) is strictly optimal at every \(A\)-node reached on \(P\). Player \(B\), by truthfulness, follows the revealed moves. Hence the resulting play follows \(P\).

The number of revelations required is at most the number of \(B\)-nodes along \(P\), which is at most \(n\).
\end{proof}

\subsection{Illustrative Example}

We illustrate the definitions and the main result with a simple monotone example, corresponding to the case $n=2$. The tree contains $16$ terminal outcomes, whose payoffs to player $A$ increase from left to right:
\[
0,1,2,\dots,15.
\]

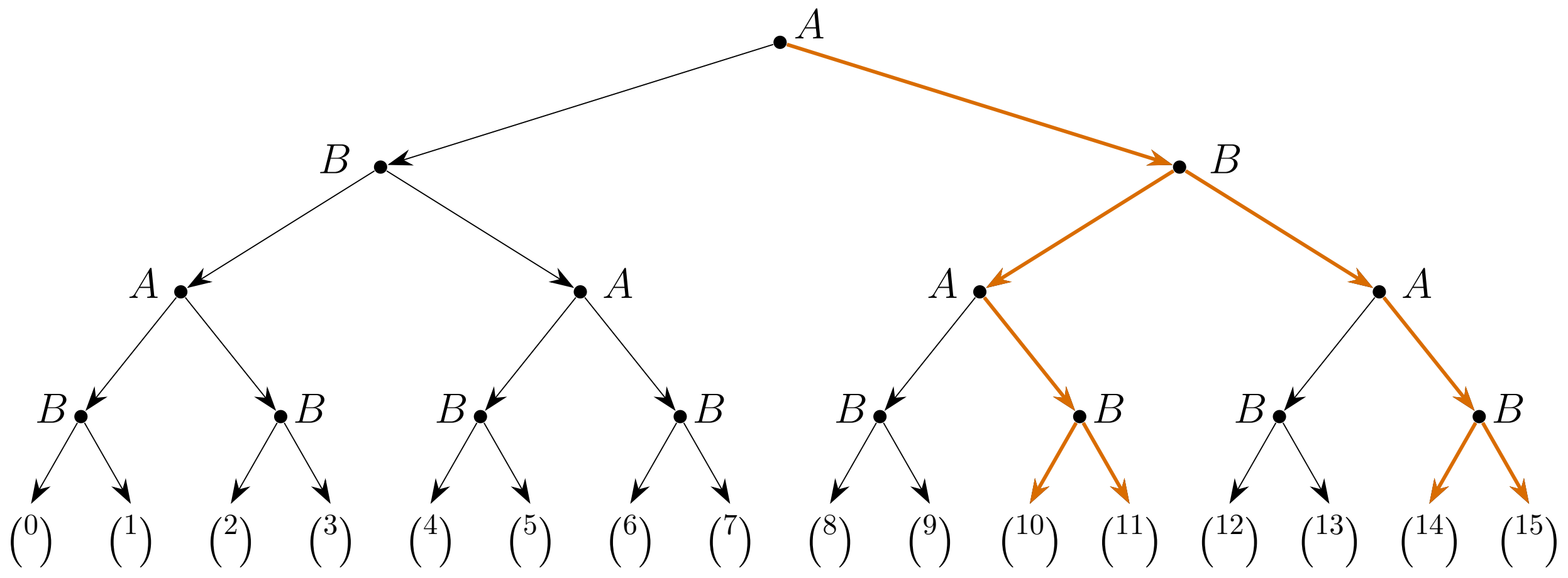
\begin{figure}[H]
\centering
\resizebox{\textwidth}{!}{%
\begin{tikzpicture}[
    x=1cm, y=1cm,
    >={Stealth[length=3mm,width=2mm]},
    every node/.style={inner sep=0pt},
    player/.style={font=\Large},
    payoff/.style={font=\Large},
    node dot/.style={circle, fill=black, inner sep=1.6pt},
    black edge/.style={->, line width=0.4pt, black},
    poss edge/.style={->, line width=1.25pt, orange!85!black}
]

\coordinate (A0) at (0,6.0);

\coordinate (B1L) at (-4.8,4.5);
\coordinate (B1R) at ( 4.8,4.5);

\coordinate (A2LL) at (-7.2,3.0);
\coordinate (A2LR) at (-2.4,3.0);
\coordinate (A2RL) at ( 2.4,3.0);
\coordinate (A2RR) at ( 7.2,3.0);

\coordinate (B3LLL) at (-8.4,1.5);
\coordinate (B3LLR) at (-6.0,1.5);
\coordinate (B3LRL) at (-3.6,1.5);
\coordinate (B3LRR) at (-1.2,1.5);
\coordinate (B3RLL) at ( 1.2,1.5);
\coordinate (B3RLR) at ( 3.6,1.5);
\coordinate (B3RRL) at ( 6.0,1.5);
\coordinate (B3RRR) at ( 8.4,1.5);

\coordinate (L1)  at (-9.0,0.0);
\coordinate (L2)  at (-7.8,0.0);
\coordinate (L3)  at (-6.6,0.0);
\coordinate (L4)  at (-5.4,0.0);
\coordinate (L5)  at (-4.2,0.0);
\coordinate (L6)  at (-3.0,0.0);
\coordinate (L7)  at (-1.8,0.0);
\coordinate (L8)  at (-0.6,0.0);
\coordinate (L9)  at ( 0.6,0.0);
\coordinate (L10) at ( 1.8,0.0);
\coordinate (L11) at ( 3.0,0.0);
\coordinate (L12) at ( 4.2,0.0);
\coordinate (L13) at ( 5.4,0.0);
\coordinate (L14) at ( 6.6,0.0);
\coordinate (L15) at ( 7.8,0.0);
\coordinate (L16) at ( 9.0,0.0);

\node[node dot] (A0dot)    at (A0)    {};
\node[node dot] (B1Ldot)   at (B1L)   {};
\node[node dot] (B1Rdot)   at (B1R)   {};
\node[node dot] (A2LLdot)  at (A2LL)  {};
\node[node dot] (A2LRdot)  at (A2LR)  {};
\node[node dot] (A2RLdot)  at (A2RL)  {};
\node[node dot] (A2RRdot)  at (A2RR)  {};
\node[node dot] (B3LLLdot) at (B3LLL) {};
\node[node dot] (B3LLRdot) at (B3LLR) {};
\node[node dot] (B3LRLdot) at (B3LRL) {};
\node[node dot] (B3LRRdot) at (B3LRR) {};
\node[node dot] (B3RLLdot) at (B3RLL) {};
\node[node dot] (B3RLRdot) at (B3RLR) {};
\node[node dot] (B3RRLdot) at (B3RRL) {};
\node[node dot] (B3RRRdot) at (B3RRR) {};

\node[player] at ($(A0)+(0.35,0.22)$) {$A$};
\node[player] at ($(B1L)+(-0.55,0.10)$) {$B$};
\node[player] at ($(B1R)+(0.55,0.10)$)  {$B$};
\node[player] at ($(A2LL)+(-0.45,0.10)$) {$A$};
\node[player] at ($(A2LR)+(0.45,0.10)$)  {$A$};
\node[player] at ($(A2RL)+(-0.45,0.10)$) {$A$};
\node[player] at ($(A2RR)+(0.45,0.10)$)  {$A$};

\node[player] at ($(B3LLL)+(-0.35,0.10)$) {$B$};
\node[player] at ($(B3LLR)+(0.35,0.10)$)  {$B$};
\node[player] at ($(B3LRL)+(-0.35,0.10)$) {$B$};
\node[player] at ($(B3LRR)+(0.35,0.10)$)  {$B$};
\node[player] at ($(B3RLL)+(-0.35,0.10)$) {$B$};
\node[player] at ($(B3RLR)+(0.35,0.10)$)  {$B$};
\node[player] at ($(B3RRL)+(-0.35,0.10)$) {$B$};
\node[player] at ($(B3RRR)+(0.35,0.10)$)  {$B$};

\node[payoff] (L1lab)  at (L1)  {$\binom{0}{\,}$};
\node[payoff] (L2lab)  at (L2)  {$\binom{1}{\,}$};
\node[payoff] (L3lab)  at (L3)  {$\binom{2}{\,}$};
\node[payoff] (L4lab)  at (L4)  {$\binom{3}{\,}$};
\node[payoff] (L5lab)  at (L5)  {$\binom{4}{\,}$};
\node[payoff] (L6lab)  at (L6)  {$\binom{5}{\,}$};
\node[payoff] (L7lab)  at (L7)  {$\binom{6}{\,}$};
\node[payoff] (L8lab)  at (L8)  {$\binom{7}{\,}$};
\node[payoff] (L9lab)  at (L9)  {$\binom{8}{\,}$};
\node[payoff] (L10lab) at (L10) {$\binom{9}{\,}$};
\node[payoff] (L11lab) at (L11) {$\binom{10}{\,}$};
\node[payoff] (L12lab) at (L12) {$\binom{11}{\,}$};
\node[payoff] (L13lab) at (L13) {$\binom{12}{\,}$};
\node[payoff] (L14lab) at (L14) {$\binom{13}{\,}$};
\node[payoff] (L15lab) at (L15) {$\binom{14}{\,}$};
\node[payoff] (L16lab) at (L16) {$\binom{15}{\,}$};

\draw[black edge] (A0dot) -- (B1Ldot);
\draw[black edge] (A0dot) -- (B1Rdot);

\draw[black edge] (B1Ldot) -- (A2LLdot);
\draw[black edge] (B1Ldot) -- (A2LRdot);
\draw[black edge] (B1Rdot) -- (A2RLdot);
\draw[black edge] (B1Rdot) -- (A2RRdot);

\draw[black edge] (A2LLdot) -- (B3LLLdot);
\draw[black edge] (A2LLdot) -- (B3LLRdot);
\draw[black edge] (A2LRdot) -- (B3LRLdot);
\draw[black edge] (A2LRdot) -- (B3LRRdot);
\draw[black edge] (A2RLdot) -- (B3RLLdot);
\draw[black edge] (A2RLdot) -- (B3RLRdot);
\draw[black edge] (A2RRdot) -- (B3RRLdot);
\draw[black edge] (A2RRdot) -- (B3RRRdot);

\draw[black edge] (B3LLLdot) -- ($(L1lab)+(0,0.45)$);
\draw[black edge] (B3LLLdot) -- ($(L2lab)+(0,0.45)$);
\draw[black edge] (B3LLRdot) -- ($(L3lab)+(0,0.45)$);
\draw[black edge] (B3LLRdot) -- ($(L4lab)+(0,0.45)$);

\draw[black edge] (B3LRLdot) -- ($(L5lab)+(0,0.45)$);
\draw[black edge] (B3LRLdot) -- ($(L6lab)+(0,0.45)$);
\draw[black edge] (B3LRRdot) -- ($(L7lab)+(0,0.45)$);
\draw[black edge] (B3LRRdot) -- ($(L8lab)+(0,0.45)$);

\draw[black edge] (B3RLLdot) -- ($(L9lab)+(0,0.45)$);
\draw[black edge] (B3RLLdot) -- ($(L10lab)+(0,0.45)$);
\draw[black edge] (B3RLRdot) -- ($(L11lab)+(0,0.45)$);
\draw[black edge] (B3RLRdot) -- ($(L12lab)+(0,0.45)$);

\draw[black edge] (B3RRLdot) -- ($(L13lab)+(0,0.45)$);
\draw[black edge] (B3RRLdot) -- ($(L14lab)+(0,0.45)$);
\draw[black edge] (B3RRRdot) -- ($(L15lab)+(0,0.45)$);
\draw[black edge] (B3RRRdot) -- ($(L16lab)+(0,0.45)$);

\draw[poss edge] (A0dot) -- (B1Rdot);

\draw[poss edge] (B1Rdot) -- (A2RLdot);
\draw[poss edge] (B1Rdot) -- (A2RRdot);

\draw[poss edge] (A2RLdot) -- (B3RLRdot);
\draw[poss edge] (A2RRdot) -- (B3RRRdot);

\draw[poss edge] (B3RLRdot) -- ($(L11lab)+(0,0.45)$);
\draw[poss edge] (B3RLRdot) -- ($(L12lab)+(0,0.45)$);
\draw[poss edge] (B3RRRdot) -- ($(L15lab)+(0,0.45)$);
\draw[poss edge] (B3RRRdot) -- ($(L16lab)+(0,0.45)$);

\end{tikzpicture}
}
\caption{Possible paths in the monotone two-level-per-player example. The bottom entries of the terminal payoffs are left blank to emphasize that possibility of paths depends only on player $A$'s payoffs and reasoning. The highlighted paths lead to the possible outcomes $10$, $11$, $14$, and $15$.}
\end{figure}

Figure~8 shows the monotone example and highlights the four possible paths.
The highlighted paths satisfy $A$'s local optimality conditions when $B$'s moves along them are revealed. 
In the absence of any revelations, player $A$ evaluates actions recursively by their guaranteed payoffs. At each $B$-node, since $A$ has no information about which move $B$ will make, both continuations are considered possible for purposes of computing $A$'s guarantee. At the root, propagating these guarantees upward gives payoff $2$ for the left subtree and $10$ for the right subtree, so $A$ selects the right branch.

Continuing this reasoning within the right subtree determines a reference path for computing $A$'s guarantee: at each $A$-node, $A$ chooses the branch with the higher guaranteed payoff, while at each $B$-node the path follows the continuation that realizes $A$'s guarantee. We refer to this baseline path as the $0$-path.

To determine which paths are possible, we evaluate each path under the convention that its moves are known to player $A$, regardless of whether any actual revelations are needed to induce this knowledge. This corresponds exactly to the definition of a possible path in Section 4.2.

\medskip

\noindent
\emph{Outcome $10$.}
This is the $0$-path. The induced guaranteed payoffs coincide with those under baseline reasoning, so the path remains optimal for $A$. Hence $10$ is possible.

\medskip

\noindent
\emph{Outcome $11$.}
This path differs from the $0$-path only in the final move of player $B$. The resulting pruning raises the target-direction guarantee from 10 to 11 while preserving $A$'s preferred directions, so the path remains optimal. Hence $11$ is possible.

\medskip

\noindent
\emph{Outcome $14$.}
This path differs from the $0$-path in the move of player $B$ at the level-1 $B$-node. The subtree that previously determined the minimum payoff at the root is eliminated, increasing the guaranteed payoff along the right branch. Recomputing at the lower $A$-node yields the same conclusion. Hence $14$ is possible.

\medskip

\noindent
\emph{Outcome $15$.}
This path further modifies the final $B$-move from the path leading to $14$. As in the case of $11$, this modification does not alter the induced preferences of player $A$, so the path remains optimal. Hence $15$ is possible.

\medskip

All other outcomes fail the condition for being possible. This occurs in two ways:

\begin{itemize}
    \item Outcomes $0$ through $7$ lie in the left subtree at the root. Even under any pruning induced by the path, the guaranteed payoff along the left branch cannot exceed that of the right branch, so player $A$ would deviate at the root.
    \item Outcomes $8,9,12,13$ lie in the right subtree but fail at the lower $A$-node. In each case, the induced minimum payoff along the path is strictly less than that of the alternative branch, so $A$ would deviate.
\end{itemize}

Thus, the only possible outcomes in this example are
\[
10, \quad 11, \quad 14, \quad 15.
\]
By Theorem 1, each of these possible paths can be implemented through truthful revelation using at most \(n=2\) disclosures.


\section{Future Work}

A natural question concerns the computational problem faced by player $B$: identifying a path that induces an outcome maximizing their payoff. While Theorem 1 guarantees that any such path can be realized using at most $n$ revelations, the number of candidate paths grows exponentially with the depth of the game tree. Identifying a payoff-maximizing possible path requires determining whether candidate paths are possible, which in turn involves evaluating the recursive reasoning they induce. This makes even checking a single path nontrivial, and developing efficient algorithms for this problem would be a natural direction for future work.

Several extensions of the model may further enrich the analysis. One direction would be to move beyond conservative (maximin) reasoning to other decision rules. Since the mechanism identified in this paper operates through the pruning of feasible outcomes, it reflects a broader phenomenon that may extend to settings where choices depend on such sets more generally. Another direction would be to allow revelations to occur dynamically during play, rather than only prior to strategy formation, introducing contingent or delayed disclosures. Extending the framework to non-generic settings, where payoff comparisons may involve ties, would also require refining the notion of possible paths and may lead to qualitatively different behavior.

Finally, it would be valuable to study richer classes of revelation strategies. In particular, off-path revelations---those not lying along the realized path---may influence behavior indirectly through their interaction with recursive reasoning. More broadly, one may ask which disclosures are beneficial or detrimental to the manipulating player in terms of the marginal contribution of individual revelations to the final outcome. These questions connect to themes in information design, where the structure and timing of information disclosure play a central role.

\section{Conclusion}

We study a model of extensive-form interaction in which strategic influence arises not from altering payoffs or feasible actions, but from shaping an opponent's perception of which outcomes remain possible. By combining conservative decision-making with endogenous information revelation, the framework captures how truthful commitments prune feasible continuations and thereby alter strategic reasoning.

Our analysis shows that this form of epistemic influence can be highly efficient. Although the underlying decision environment may contain exponentially many outcomes, inducing a desired outcome requires only a number of revelations that grows linearly with the depth of the game. This reflects a fundamental contrast between the combinatorial complexity of the environment and the limited informational effort needed to guide its effective resolution.

Central to this perspective is the characterization of outcomes in terms of possible paths: paths that remain optimal for player A when their moves are treated as known. Our main result shows that every such path can be implemented by player B through truthful revelation, providing a structural link between epistemic reasoning and strategic control.

More broadly, these results highlight how decision-making based on feasible outcome sets and epistemic structure interact in dynamic settings. Rather than treating uncertainty as exogenous or resolving it through fixed decision rules, the model shows how knowledge, reasoning, and feasible continuations co-evolve through disclosure. This provides a framework for understanding how minimal but well-targeted informational interventions can exert disproportionate strategic influence in complex decision environments.

\bibliographystyle{plainnat}
\bibliography{references}

\end{document}